\documentclass[%
 reprint,
nofootinbib,
nobibnotes,
 amsmath,amssymb,
 aps, prd,
 longbibliography,
floatfix,
]{revtex4-2}

\usepackage{graphicx}% Include figure files
\usepackage{dcolumn}% Align table columns on decimal point
\usepackage{bm}% bold math
\usepackage[normalem]{ulem}
\usepackage[english]{babel}
\usepackage{fancyhdr}
\usepackage{amsmath}
\usepackage{amssymb}
\usepackage{amsfonts}
\usepackage{psfrag}
\usepackage[utf8]{inputenc}
\usepackage[bf,footnotesize,justification=Justified,format=plain]{caption}
\usepackage[dvipsnames]{xcolor}
\usepackage[colorlinks=True, citecolor=blue, linkcolor=blue, urlcolor=blue,linktocpage]{hyperref}
\usepackage{amsthm}
\usepackage{gensymb}
\usepackage{subfig}
\usepackage{physics}
\usepackage{array}
\usepackage{tcolorbox, mathtools}
\usepackage{soul}
\tcbuselibrary{skins}
\usepackage{aas_macros}
\usepackage{cleveref}
\usepackage{orcidlink}

\newcommand{\centra}{CENTRA, Departamento de Física, Instituto Superior Técnico – IST,
Universidade de Lisboa – UL, Avenida Rovisco Pais 1, 1049-001 Lisboa, Portugal}
\newcommand{\dfa}{Dipartimento di Fisica e Astronomia ‘G. Galilei’, Università di Padova, Via Francesco Marzolo 8, 35131 Padova, Italy}
\newcommand{\infnpd}{INFN Sezione di Padova, Via Francesco Marzolo 8, 35131 Padova, Italy}
\newcommand{\sissa}{SISSA, Via Bonomea 265, 34136 Trieste, Italy \& INFN Sezione di Trieste}
\newcommand{\ifpu}{IFPU - Institute for Fundamental Physics of the Universe, Via Beirut 2, 34014 Trieste, Italy}

\newcommand{\la}{\langle}
\newcommand{\ra}{\rangle}

\begin{document}

\title{Looking for non-gaussianity in Pulsar Timing Arrays through the four point correlator}% Force line breaks with \\
%Towards implementing non-gaussianity in the Pulsar Timing Array pipeline: the role of the four point function.
%Looking for non-gaussianity in Pulsar Timing Arrays through the four point function

\author{Adrien Kuntz~\orcidlink{0000-0002-4803-2998}}
\email{adrien.kuntz@tecnico.ulisboa.pt}
\affiliation{\centra}

\author{Clemente Smarra~\orcidlink{0000-0002-0817-2830}}
\email{clemente.smarra@unipd.it}
\affiliation{\dfa}
\affiliation{\infnpd}

\author{Massimo Vaglio~\orcidlink{0000-0002-7285-3489}}
\email{mvaglio@sissa.it}
\affiliation{\sissa}
\affiliation{\ifpu}

\begin{abstract}
\noindent
Pulsar Timing Arrays have recently reported strong evidence for a stochastic gravitational wave background. In standard analyses, it is modeled through pulsar-dependent Fourier coefficients assumed to follow gaussian statistics, so that the signal is fully characterized by its two-point function. However, if the background arises from a finite population of inspiralling supermassive black hole binaries, non-gaussian features may emerge, making the determination of higher-order correlators essential. 
\noindent
In this work, we compute the complete four-point correlator of the stochastic gravitational wave background Fourier coefficients for four arbitrary pulsar positions, identifying it as the leading probe of non-gaussianity. The result separates into a gaussian contribution, proportional to the square of the two-point function, and a genuinely non-gaussian connected component, whose non-trivial angular dependence generalizes the Hellings and Downs correlation to four pulsars.
This angular structure depends only on averages of products of antenna pattern functions, and is therefore expected to be independent of the specific physical origin of the background.
We further propose to incorporate the four-point correlator into the parameter-estimation pipeline by deriving a marginalized likelihood that perturbatively accounts for non-gaussian effects. 
Our results provide the theoretical framework to search for non-gaussian features in pulsar timing array data, opening the way to a more complete characterization of gravitational-wave backgrounds.
\end{abstract}

\maketitle

\section{Introduction}
In recent years, Pulsar Timing Array (PTA) collaborations reported evidence for a temporally-correlated stochastic process leaving its imprints on the times of arrival (TOAs) of radio beams coming from a network of monitored millisecond pulsars in our Galaxy~\cite{EPTAIII, Agazie_2023_SGWB, Reardon_2023, Xu_2023, Miles_2024}. 
This process exhibits inter-pulsar correlations consistent with the Hellings and Downs (HD) curve~\cite{1983ApJ...265L..39H}
\begin{equation} \label{eq:HD}
    \mu(\gamma_{ab}) = \frac{1}{4} + \frac{\cos \gamma_{ab}}{12} + \frac{1-\cos \gamma_{ab}}{2} \log \bigg(\frac{1-\cos \gamma_{ab}}{2} \bigg),
\end{equation}
which is a fundamental prediction of General Relativity for the angular correlation pattern induced by gravitational waves (GWs) between two pulsars $a$, $b$, separated by an angle $\gamma_{ab}$. These observations point to the existence of a stochastic gravitational wave background (SGWB) permeating the Universe, whose origin is however still debated. The current leading interpretation is that the SGWB arises from an assembly of inspiralling supermassive black holes binaries (SMBHBs) emitting GWs via the quadrupole formula~\cite{Rajagopal:1994zj, Phinney:2001, Jaffe:2002rt,Wyithe_2003, Sesana_2008, Burke-Spolaor:2018bvk}, although early universe physics processes can also be invoked~\cite{Sanidas_2012,Caprini:2018mtu, Blasi_2021, Chen_2022,   EPTA_IV, Afzal_2023, Figueroa:2023zhu,  Madge:2023dxc,   Franciolini:2023wjm, Franciolini:2023pbf}.  In this work, we assume that the SGWB is purely of astrophysical origin, thus produced by a large number of inspiralling SMBHBs. However, as we will highlight in the following, our main result remains valid regardless of the specific origin of the background, provided it consists of GWs.\\

PTA analyses typically assume that the SGWB is gaussian, implying that it is fully characterized by its two-point (2PT) correlation function. By the central limit theorem, this assumption holds well when the SGWB is generated by a large number of uncorrelated sources~\cite{Allen:2022dzg,Allen:2024rqk,Allen:2022ksj,Allen:2024uqs,Caprini:2018mtu}. In this case, cross-correlating the TOAs of pulsar pairs suffices to determine the properties of the background: the correlation amplitude encodes the SGWB power spectrum, while the HD curve captures the angular dependence.\\

While gaussianity is a powerful and often justified assumption, it may break down for an astrophysical signal produced by SMBHBs. In fact, because quadrupole emission causes binaries to evolve slowly at low frequencies and rapidly at higher frequencies, the number of contributing sources depends strongly on the observing GW frequency~\cite{Sesana_2008}. Consequently, while PTA observations at low frequencies are deep in the gaussian regime, higher frequencies are expected to be dominated by a handful of sources, and the central limit theorem does not apply.\\

This implies that an astrophysical background can exhibit measurable non-gaussian features, depending on the frequency range considered, and neglecting them may bias the inferred signal properties.
Such non-gaussianity immediately translates in non-trivial higher order correlators that cannot be entirely described in terms of the power spectrum and the HD angular pattern~\cite{Xue:2024qtx,Bernardo:2024uiq}. 

A natural question thus arises: what are the expected amplitude and angular structure of higher-order correlators of an astrophysical SGWB?
A first possibility is to consider the three-point (3PT) correlator, or bispectrum~\cite{Powell:2019kid,Tsuneto:2018tif,Adshead:2009bz}. However, under our basic assumption that the SGWB is produced by a large number of independent sources, the bispectrum is expected to vanish on average. This follows from a simple property of random phases: the product of three independent random and uniformly distributed phases averages to zero. 
Here, by ``average" we mean an ensemble average over random directions and phases of the SMBHBs, which is physically justified since there are many sources.
%\CS{I still find the sentence confusing to a person who reads this for the first time. I would remove the part ``which is .... But up to you.}
%\AK{\sout{performing the sum over the large number of sources of GW \CS{Although I see the point, I don't like this formulation. it's more an ensemble average that we are doing in the following,no? I mean, according to our computation in the following, we would have 0 even if we have just 3 BHs (or even 1!). What you are saying here is true (and to some extent more powerful), but different than what we do. If you agree, I'll modify this sentence accordingly.}. \AK{Let's discuss it next call} \CS{I always fear these sentences.}}}
Hence, it can be argued that the bispectrum is suppressed in any scenario where the random phase argument applies~\cite{Kehagias:2024plp, Bartolo:2018evs, Margalit:2020sxp, Bartolo:2018rku}.  Note that, although the mean of the 3PT correlator is expected to vanish, its variance can be nontrivial. This means that in a given universe realization we could still be able to observe a nonzero 3PT correlator, see e.g. Ref.~\cite{Srednicki:1993ix} for a similar feature in CMB observations.

We are thus naturally led to consider the four-point (4PT) function as the lowest-order non-trivial correlator sensitive to non-gaussianity. In this case, the random phase argument is no longer applicable, as one can correlate four contributions arising from the \textit{same} SMBH source. In fact, by constructing an appropriate observable, the phase dependence can cancel out, yielding a non-zero excess kurtosis. This argument has been exploited in Ref.~\cite{Lamb:2025niq}, where the amplitude of the 4PT function for a single pulsar correlated with itself four times was estimated using astrophysical population models. As anticipated, the resulting excess kurtosis goes to zero in the limit of a large number of SMBHBs emitting GWs. \\

However, that work did not determine the full angular dependence of the 4PT function as a function of four distinct pulsar positions. This information is essential for searching for non-gaussianity in real PTA data. Indeed, at the level of the 2PT function, the shape of the HD function is crucial for distinguishing what is actually a SGWB from a common red noise. By analogy, any search for non-gaussian features will need the equivalent of the HD function for four pulsars. In this article, we hence derive the complete angular correlation of the 4PT function, which consitutes our main result.\\
\newline
We then show how the 4PT correlator enters the PTA parameter estimation pipeline. This issue was recently addressed in Ref.~\cite{Falxa:2025}, where the gaussian assumption was improved upon by using a gaussian mixture model. While that work carefully demonstrated how non-gaussianity modifies the marginalized likelihood, it did not compute the 4PT explicitly, leaving the result expressed in terms of phenomenological non-gaussian parameters. 
\newline
In this work, we aim at filling these two gaps. First, we derive the complete functional form of the 4PT correlator in terms of four generic pulsars positions. Second, we consistently incorporate the 4PT in the PTA pipeline, providing an explicit expression for the marginalized likelihood that may be implemented in the PTA data analysis routine.  \\
\newline
The outline of our work is as follows. In Sec.~\ref{sec:bisp}, we introduce a toy model of a population of SMBHBs generating the SGWB, along the lines of Ref.~\cite{Allen:2022dzg}, which we refer to for further details. Within this setup, we review the derivation of the 2PT function, making use of relations from Ref.~\cite{Allen:2022dzg}. In particular, we show that the average of the 2PT function over different realizations of the SMBHB population is proportional to the HD correlation in Eq.~\eqref{eq:HD}, and we explicitly demonstrate why the 3PT function vanishes on average. \\
In Sec.~\ref{sec:4pt}, we then turn to the 4PT correlator, which represents the leading non-gaussian contribution. We show that its average consists of two parts: a gaussian component, proportional to the square of the 2PT correlator, and a genuinely non-gaussian term, namely the \textit{connected} 4PT correlator. In particular, we compute the complete angular dependence of the 4PT function in the most general configuration, correlating four pulsars.
While our derivation relies on the toy model detailed in Sec.~\ref{sec:bisp}, we argue that our result is valid in a broader context. Indeed, analogously to the HD for the 2PT, the angular dependence of the 4PT is only determined by averages of products of antenna pattern functions, and it is thus insensitive on the underlying physical mechanism.\\
Having established the form of the 4PT correlator, we describe how it enters the PTA pipeline in Sec.~\ref{sec:pta}. We begin in Sec.~\ref{subsec:gauss} with a brief review of the noise sources present in PTA data and of the modeling of a SGWB in terms of Fourier coefficients, finally presenting the standard gaussian marginalized likelihood used to perform statistical inference on a gaussian SGWB. 
Then, in Sec.~\ref{subsec:nglik}, we extend this framework to incorporate the leading non-gaussian contributions, encoded in the 4PT derived in Sec.~\ref{sec:4pt}.
Finally, we summarize our results and present our conclusions in Sec.~\ref{sec:concl}.
Several technical details are collected in a series of appendices.

\section{Review of the two-point correlator}\label{sec:bisp}
As we stated in the introduction, we assume that the observed SGWB is generated by SMBHBs. In particular, building on Refs.~\cite{Allen:2022dzg, Allen:2022ksj}, we make the simplifying assumption that all the SMBHBs are observed face-on in circular orbits. In this way, we can group the ensemble of SMBHBs in frequency bins, according to their emitted GW frequency. Accounting for a distribution of binary inclinations is expected to affect primarily the
overall normalization of the background, while leaving its spectral and
angular structure unchanged \cite{Ali-Haimoud:2026sbk}. We also assume that all the SMBHBs in the $I$-th frequency bin are effectively characterized by the same frequency, which we identify by $\omega_I = I\cdot2\pi/T$, where $T$ is the duration of the considered PTA experiment. Thus, the deviations in the TOAs of radio beams from a given pulsar $a$ induced by the SMBHB ensemble -- commonly referred to as timing residuals~\footnote{We will be more precise about the definition of timing residuals in Sec.~\ref{sec:pta}.}-- can be written as~\cite{Allen:2022dzg}:
\begin{align} \label{eq:timingResidualSGWB}
    \delta t^a(t) = &\sum_{I=1}^{N_f} \sum_i^{N_I} c_{I,i}^a \, \text{e}^{\mathrm{i}(\omega_I t + \phi_{I,i})} + \text{c.c.}\\
    c_{I,i}^a = &\frac{A_I^i}{2 \mathrm{i} \omega_I} \left[ 1 - \text{e}^{- \mathrm{i} \omega_I L_a(1 + \vec{p}_a\cdot\vec{\Omega}_{I,i})}\right] \times\nonumber\\
    &\times \left(F_a^+(\vec\Omega_{I,i}) - \mathrm{i} F_a^\times (\vec\Omega_{I,i})\right) \label{eq:defcI}.
\end{align}
In this equation, the index $i$ labels the GW source and $N_I$ is the number of SMBHBs in the $I$-th frequency bin. Notice that we denote by $\mathrm{i}$ the complex number such that $\mathrm{i}^2=-1$ to avoid confusion with the SMBHB index $i$.
As a zero-frequency term corresponds to an unobservable shift of timing residuals,~\footnote{Technically, the timing model fit (see Sec.~\ref{sec:pta}) removes any constant offset present in the timing residuals, thus centering them on zero. More generally it suppresses power at the lowest frequencies, including sub-nHz modes \cite{Hazboun:2019vhv, Pitrou:2024scp}, and can induce nonstationarity in the post-fit residuals \cite{Allen:2025waa}.} the frequency bin index runs from $I = 1$ up to an arbitrary $N_f$, which is typically chosen to be $N_f=14$ in SGWB searches~\cite{Afzal_2023, Agazie_2023_SGWB}.
The $\phi_{I,i}$ are random phases associated with each SMBHB, $L_a$ is the distance of the pulsar $a$ from Earth, $\vec p_a$ is the unit vector pointing to pulsar $a$, and $\vec \Omega_{I,i}$ is the unit vector  identifying the direction of the $i$-th SMBHB in the $I$-th frequency bin. The oscillating term $\text{e}^{- \mathrm{i}\omega_I L_a(1 + \vec{p}_a\cdot\vec{\Omega}_{I,i})}$ is commonly referred to as \textit{pulsar term}.
Finally, $A_I^i$ is the GW amplitude, which can be related to the chirp mass and frequency of the SMBHB using the quadrupole formula, and $F_a^A$ ($A=+,\times$) are the usual angular antenna pattern functions
\begin{align} \label{eq:defFa}
    F_a^A(\vec \Omega) &= \frac{1}{2} \frac{e_{ij}^A(\vec \Omega) p_a^i p_a^j}{1+\vec p_a \cdot \vec \Omega} \; , \\
    e_{ij}^+(\vec \Omega) &= m_i m_j - n_i n_j \; , \quad e_{ij}^\times(\vec \Omega) = m_i n_j + n_i m_j \; ,
\end{align}
the two vectors $\vec m$ and $\vec n$ forming an orthonormal basis together with $\vec\Omega$.
There is an ambiguity in the definition of $\vec m$ and $\vec n$ since they can be rotated by a polarization angle $\psi$, under which we have the transformation $F_a^+ - \mathrm{i} F_a^\times \rightarrow e^{-2 \mathrm{i} \psi} (F_a^+ - \mathrm{i} F_a^\times)$. Our final result for the angular dependence of the 4PT function presented in Sec.~\ref{sec:4pt} will be independent of $\psi$, as expected (see Appendix~\ref{app:lambdaabcd}). 

By defining
\begin{align}\label{eq:Ccoeff}
    \mathcal{C}_I^a = \sum_i c_{I,i}^a e^{\mathrm{i} \phi_{I,i}} \qquad 
     \mathcal{C}_{-I}^a =  (\mathcal{C}_{I}^a)^* ,
\end{align}
Eq.~\eqref{eq:timingResidualSGWB} can be rewritten as 
\begin{equation}\label{eq:complFour}
    \delta t^a(t) = \sum_I \mathcal{C}_I^a ~e^{\mathrm{i}\omega_I t},
\end{equation}
where the index $I$ runs from $-N_f$ to $N_f$, excluding $I=0$. Eq.~\eqref{eq:complFour} shows that the timing residuals induced by the SMBHB ensemble on pulsar $a$ can be rewritten in terms of a \textit{complex} Fourier series, where $\mathcal{C}_I^a$ are the \textit{complex} Fourier coefficients and $\exp(\mathrm{i}\omega_I t)$ are the \textit{complex} Fourier basis functions. We will draw a more detailed connection between the statistics of the Fourier coefficients and the SGWB power spectral density in Sec.~\ref{sec:pta}.\\
In PTAs, we are usually interested in correlations among timing residuals. From Eq.~\eqref{eq:complFour}, it is then clear that we should find the statistical distribution of the Fourier coefficients. %This can be obtained by averaging the relevant quantities over the random phases $\phi_{I,i}$ and the sky positions $\vec \Omega_{I,i}$ of the SMBHBs. 
This can be obtained by averaging any relevant quantity of interest (say $\mathcal{A}$) over the random phases $\phi_{I,i}$ and the sky positions $\vec \Omega_{I,i}$ of the SMBHBs:
\begin{equation}
    \langle \mathcal{A} \rangle_{\phi, \vec \Omega} = \frac{1}{8 \pi^2} \int \mathrm{d} \phi \, \mathrm{d}^2 \Omega \, \mathcal{A}.
\end{equation}
This average is justified when the number of sources is large enough, which could be an incorrect approximation for the highest frequency bins of PTAs~\cite{Sesana_2008}. For a more detailed discussion, see Sec.~\ref{subsec:nglik}.
For convenience, we first perform the average over $\phi_{I,i}$, and then the one over $\vec{\Omega}_{I,i}$. Looking at Eq.~\eqref{eq:Ccoeff}, we immediately find that $\la\mathcal{C}_I^a\ra_\phi = 0$.
Similarly $\langle \mathcal{C}_I^a  \mathcal{C}_J^b \rangle_\phi = 0 $ if $I$ and $J$ are of the same sign, because $\langle e^{\mathrm{i}( \phi_{I,i} + \phi_{J,j})} \rangle_\phi = 0$.
On the other hand, for $I,J>0$, since we have $\langle e^{\mathrm{i}( \phi_{I,i} - \phi_{J,j})} \rangle_\phi = \delta_{IJ} \delta_{ij}$,
\begin{equation}
    \langle \mathcal{C}_I^a  \mathcal{C}_{-J}^b \rangle_\phi = \delta_{IJ} \sum_i c_{I,i}^a (c_{I,i}^b)^*.
\end{equation}
Using formulae from Ref.~\cite{Allen:2022dzg}, we can now compute the mean over the source angle $\vec \Omega$:
\begin{equation} \label{eq:2ptCorr}
    \langle \mathcal{C}_I^a  \mathcal{C}_{-J}^b \rangle_{\phi,\vec \Omega} = \frac{1}{4 \omega_I^2}\delta_{IJ} \sum_i (A_I^i)^2 \bar \mu(\gamma_{ab} ),
\end{equation}
where $\gamma_{ab}$ is the angle between pulsars $a$ and $b$, and $\bar \mu(\gamma)$ is given in terms of the HD correlation function $\mu(\gamma)$ introduced in Eq.~\eqref{eq:HD} as
\begin{equation}
    \bar \mu(\gamma) =  \mu(\gamma_{ab}) +  \delta_{ab} \mu(0) . 
\end{equation}
The $\delta_{ab}  \mu(0)$ term comes from the pulsar term: when pulsar $a$ is different from pulsar $b$, the oscillating term  $\text{e}^{-\mathrm{i}\omega_I L_a(1 + \vec{p}_a\cdot\vec{\Omega}_{I,i})}$ just averages to zero ($\omega_I L_a \gg 1$ for typical pulsar distances), while it gives a factor of $2$ otherwise.~\footnote{Notice that a pulsar term may also be present for
$a\neq b$ if the pulsar-term phases remain coherent between the two pulsars. In practice, this requires not only that the pulsars be separated by less than a GW wavelength, but also that their distance uncertainties be smaller than the characteristic GW wavelength (typically a few parsecs in the PTA band) \cite{2025arXiv251210795T}. While this is not relevant for current datasets, it may become important in the future.} We note that the HD normalization used here follows the original convention of
Hellings and Downs~\cite{1983ApJ...265L..39H}, whereas modern PTA analyses often adopt a normalization in
which Earth and pulsar terms contribute equally to the total power~\cite{Romano:2023zhb}.
Both conventions differ only by an overall rescaling.
This completes the derivation of the 2PT function of the Fourier coefficients. In PTA data, this result is used to model the prior on $ \mathcal{C}_I^a $ as a gaussian with variance proportional to the HD correlation function (see Sec.~\ref{sec:pta} for more details).
\\
In this work, we seek going beyond the usual 2PT correlator. As the phases $\phi_{I,i}$ associated with the SMBHBs are random, Eq.~\eqref{eq:Ccoeff} immediately shows that the average of the three point correlator is vanishing, namely $\la\mathcal{C}_I^a \mathcal{C}_J^b \mathcal{C}_K^c\ra = 0$.
For the same reason, the average of all odd correlators vanishes, and we expect a symmetric probability distribution for $C_I^a$. \\
Therefore, the lowest non-trivial correlator we should be interested in is the 4PT function, whose computation represents one of the main results of this article and is detailed in Sec.~\ref{sec:4pt}. 

At this point, a comment is in order. Ref.~\cite{Allen:2022dzg} makes clear that the HD in Eq.~\eqref{eq:HD} represents the \textit{expectation value} of the 2PT correlator defined in Eq.~\eqref{eq:2ptCorr}. In any specific realization of the Universe, however, the measured 2PT correlation need not coincide exactly with the HD prediction, and the deviations are quantified by the variance of the 2PT correlator, which is contained in the 4PT correlator itself.\\

The same consideration applies to higher order statistics. Although the 3PT correlator (and, in general, any odd correlator) vanishes in the ensemble average, this does not imply that it must vanish in a particular realization. Indeed, the variance of the 3PT correlator is generally non-zero. However, this variance is formally contained in the 6PT correlator. Consequently, it enters at higher order than the 4PT correlator considered in this work, and can therefore be consistently neglected at our level of approximation.

\section{The four-point correlator}\label{sec:4pt}
In this Section, we compute the 4PT correlator for the model described in Sec.~\ref{sec:bisp}. We will see that the 4PT correlator consists of a gaussian component, proportional to the square of the 2PT function, and a genuinely non-gaussian contribution. While the model considered is highly simplified and can provide only an order-of-magnitude estimate of the amplitude of the non-gaussian component, it nevertheless allows us to determine the full angular dependence of the 4PT correlator, which constitutes the main result of this work. 

In analogy with the HD curve, which is universal and independent of the specific origin of the SGWB, our result may be applicable to a broad class of non-gaussian SGWB. Indeed, the angular dependence is only determined by averages of products of antenna pattern functions, and is therefore insensitive to the underlying physical mechanism sourcing the SGWB. This result holds within the approximation where the oscillatory, frequency-dependent pulsar terms are neglected, yielding a purely geometric structure analogous to the HD curve. 
\\ Conversely, the overall amplitude of the 4PT correlator may carry information about the physical source of non-gaussianity.

We now want to find the higher-order moments of $ \mathcal{C}_I^a $. As the average of three $ \mathcal{C}_I^a $ vanishes, the lowest non-trivial correlator that we have to compute is the 4PT function. Once again, taking $I,J,K,L>0$, the only non-trivial moment (modulo permutations) is: 
\begin{align}
    \mathcal{C}_I^a  \mathcal{C}_{-J}^b  \mathcal{C}_K^c  \mathcal{C}_{-L}^d &= \sum_{i,j,k,l} c_{I,i}^a (c_{J,j}^b)^* c_{K,k}^c (c_{L,l}^d)^* \times \nonumber \\
    &\times \exp \mathrm{i}(  \phi_{I,i} - \phi_{J,j} +  \phi_{K,k} - \phi_{L,l} ).
\end{align}
We average over the angles $\phi$ to find
\begin{align}\label{eq:fullcorr}
    &\langle \mathcal{C}_I^a  \mathcal{C}_{-J}^b  \mathcal{C}_K^c  \mathcal{C}_{-L}^d \rangle_\phi  = \delta_{IJ} \delta_{KL} \delta_{IK} \bigg( \sum_i c_{I,i}^a  (c_{I,i}^b)^* c_{I,i}^c (c_{I,i}^d)^*  \nonumber \\
    &+\sum_{i \neq k} c_{I,i}^a  (c_{I,i}^b)^* c_{I,k}^c (c_{I,k}^d)^* + c_{I,i}^a  (c_{I,i}^d)^* c_{I,k}^c (c_{I,k}^b)^* \bigg) \nonumber \\
    &+ \delta_{IJ} \delta_{KL} (1-\delta_{IK}) \bigg(\sum_{i, k} c_{I,i}^a  (c_{I,i}^b)^* c_{K,k}^c (c_{K,k}^d)^* \bigg) \nonumber \\
    &+ \delta_{IL} \delta_{JK} (1-\delta_{IK}) \bigg( \sum_{i, k}  c_{I,i}^a  (c_{I,i}^d)^* c_{K,k}^c (c_{K,k}^b)^* \bigg), 
\end{align}
where the $i \neq k$ index in the second sum indicates that we are considering different SMBHBs. 
The first two lines of this expression account for all SMBHBs within the same frequency bin, whereas the last two lines involve contributions from two distinct frequency bins. Notice that, in the latter case, the SMBHB indices $i$ and $k$ necessarily correspond to different binaries , since they are located in different frequency bins. 
We now perform the average over the angle $\vec \Omega$. Let us denote $\mathcal{H}_2^I = \sum_i (A_I^i/\omega_I)^2$ and $\mathcal{H}_4^I = \sum_i (A_I^i/\omega_I)^4$ (notice that $\mathcal{H}_2^I$ and $\mathcal{H}_4^I$ differ from the expressions in Ref.~\cite{Allen:2022dzg} by factors of $\omega_I^{-1}$, as we are looking at timing residuals instead of redshifts). The first sum is the most involved, as it contains an average involving all four pulsars simultaneously. On general grounds, it can be expressed as:
\begin{equation} \label{eq:deflambdaabcd}
    \sum_i \langle c_{I,i}^a  (c_{I,i}^b)^* c_{I,i}^c (c_{I,i}^d)^* \rangle_
    {\vec\Omega} = \frac{1}{16} \mathcal{H}_4^I \eta_{abcd} \lambda_{abcd},
\end{equation}
where the angular factor $\lambda_{abcd} \sim \mathcal{O}(1)$ depends on the positions of all the four pulsars $a,b,c,d$. The factor $\eta_{abcd}$ accounts for the pulsar term contributions in the average and is given by
\begin{equation}
    \eta_{abcd} = \begin{cases}
        6 \, , \, a=b=c=d \\
        4 \, , \, (a=b \, \text{and} \, c=d) \, \text{or} \, (a=d \, \text{and} \, b=c) \\
        2 \, , \, a=b \, \text{or} \, c=d \, \text{or} \, a=d \, \text{or} \, b=c \\
        1 \, , \, \text{otherwise},
    \end{cases}
\end{equation}
as can be found from Eq.~\eqref{eq:defcI}.
The calculation of $\lambda_{abcd}$ is detailed in Appendix~\ref{app:lambdaabcd}, while its functional dependence on the pulsars angles is discussed in Section~\ref{sec:connected4PT}. 
The remaining terms in Eq.~\eqref{eq:fullcorr} are straightforward to compute, because they involve averages over only two pulsars at a time. These can easily be computed using the same formulae employed for Eq.~\eqref{eq:2ptCorr}. We finally get
\begin{align} \label{eq:4ptCorr}
    &16 \langle \mathcal{C}_I^a  \mathcal{C}_{-J}^b  \mathcal{C}_K^c  \mathcal{C}_{-L}^d \rangle_{\phi, \vec\Omega}  = \delta_{IJ} \delta_{KL} \delta_{IK} \mathcal{H}_4^I \bigg( \eta_{abcd}  \lambda_{abcd} \nonumber \\
    &- \big[ \bar \mu(\gamma_{ab}) \bar \mu(\gamma_{cd}) + \bar \mu(\gamma_{ad}) \bar \mu(\gamma_{bc}) \big]  \bigg)  \nonumber \\ 
    &+ \mathcal{H}_2^I \mathcal{H}_2^K  \bigg[ \delta_{IJ} \delta_{KL}  \bar \mu(\gamma_{ab}) \bar \mu(\gamma_{cd}) + \delta_{IL} \delta_{JK} \bar \mu(\gamma_{ad}) \bar \mu(\gamma_{bc})    \bigg],
\end{align}
where we used $\sum_{i \neq k} (A_I^i/\omega_I)^2 (A_I^k/\omega_I)^2 = (\mathcal{H}_2^I)^2 - \mathcal{H}_4^I$.

It turns out that the last line of this expression just consists of the usual 4PT function of a gaussian field. Indeed, using Wick's theorem~\cite{Wick:1950} for a gaussian field, 
\begin{align}
    &\langle \mathcal{C}_I^a  \mathcal{C}_{-J}^b  \mathcal{C}_K^c  \mathcal{C}_{-L}^d \rangle_{\mathrm{Gaussian}}  = \langle  \mathcal{C}_I^a  \mathcal{C}_{-J}^b \rangle \langle  \mathcal{C}_K^c  \mathcal{C}_{-L}^d \rangle \nonumber \\
    &+ \langle  \mathcal{C}_I^a  \mathcal{C}_{K}^c \rangle \langle  \mathcal{C}_{-J}^b  \mathcal{C}_{-L}^d \rangle + \langle  \mathcal{C}_I^a  \mathcal{C}_{-L}^d \rangle \langle  \mathcal{C}_{-J}^b  \mathcal{C}_{K}^c \rangle,\label{eq:wick}
\end{align}
which reproduces the last line of Eq.~\eqref{eq:4ptCorr} (indeed, notice that the second term in Eq.~\eqref{eq:wick} vanishes for our particular choice of indices). Hence, the non-gaussian part of the coefficient $\mathcal{C}_I^a$ consists of the first two lines of Eq.~\eqref{eq:4ptCorr}, namely the \textit{connected} 4PT correlation function. This term is proportional to $\mathcal{H}_4^I$, and appears only when all fields belong to the same frequency bin. Using the results of Ref.~\cite{Allen:2022dzg}, we can see that if a frequency bin contains a large number of sources, then $\mathcal{H}_4^I \ll (\mathcal{H}_2^I)^2$, implying that the resulting non-gaussianity is indeed small. See also Ref.~\cite{Lamb:2025niq} for more refined results on the scaling of the amplitude of non-gaussianities using more realistic astrophysical models.

In the next Section, we  discuss the properties of $\lambda_{abcd}$, which is the non-trivial part of the non-gaussianity of $\mathcal{C}_I^a$.

\subsection{The connected four-point correlator} \label{sec:connected4PT}

The function $\lambda_{abcd}$ depends on the relative angles between all pulsars, which determine their spatial configuration. In a given coordinate system, any four directions $\vec{p}_i \in S^2$ ($i=a,b,c,d$) are specified by $2\times 4=8$ parameters. However, the whole configuration is left unchanged under global $SO(3)$ rotations which can be used to fix three of the eight degrees of freedom.

\begin{figure}
    \centering
    \includegraphics[width=1\linewidth]{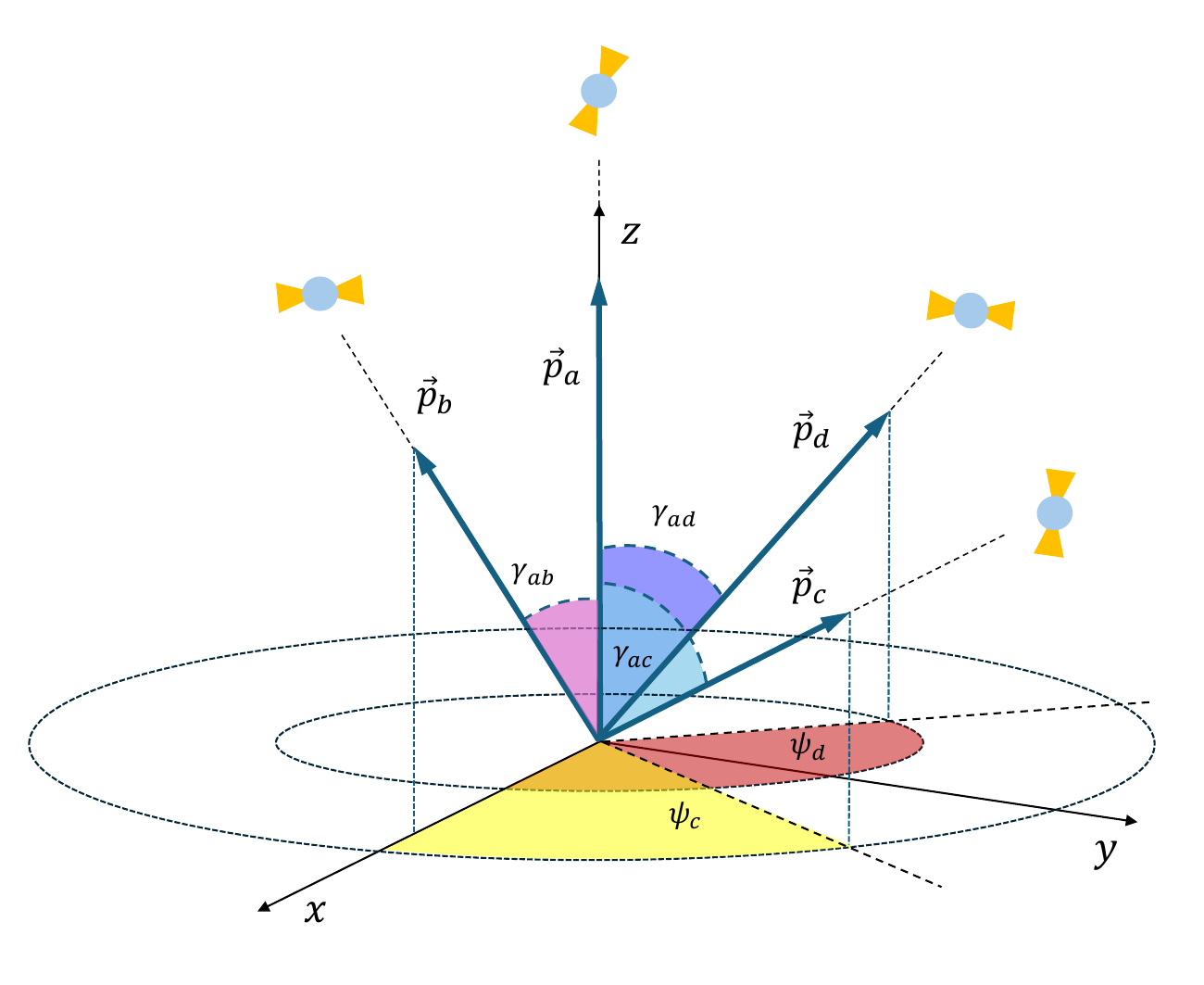}
    \caption{Schematic illustration of the chosen coordinate system for evaluating $\lambda_{abcd}$. Pulsar $a$ is aligned along the $z$-axis, pulsar $b$ lies in the $xz$ plane, forming an angle $\gamma_{ab}$ with pulsar $a$, and pulsars $c$ and $d$ are parametrized by their polar angles $\gamma_{ac}$, $\gamma_{ad}$ and azimuthal angles $\Psi_c$, $\Psi_d$, respectively.
    }
    \label{fig:pulsar_conf}
\end{figure}

In evaluating $\lambda_{abcd}$ we choose a coordinate system in which pulsar $a$ lies on the $z$ direction and pulsar $b$ is on the $xz$ plane, as illustrated in Fig.~\ref{fig:pulsar_conf}. Hence, we parametrize the pulsars unit vectors as
\begin{align} \label{eq:coordinateSystem}
    \vec p_a &= \begin{pmatrix}
        0 \\ 0 \\ 1
    \end{pmatrix} \; , \quad 
    \vec p_b = \begin{pmatrix}
        \sin \gamma_{ab} \\ 0 \\  \cos \gamma_{ab}
    \end{pmatrix} \; , \\
    \vec p_c &= \begin{pmatrix}
        \sin \gamma_{ac} \cos \Psi_c \\ \sin \gamma_{ac} \sin \Psi_c  \\  \cos \gamma_{ac}
    \end{pmatrix} \; , \quad 
    \vec p_d = \begin{pmatrix}
        \sin \gamma_{ad} \cos \Psi_d \\ \sin \gamma_{ad} \sin \Psi_d  \\  \cos \gamma_{ad}
    \end{pmatrix} \nonumber \; ,
\end{align}
and $\lambda_{abcd}$ will depend on the five angles $\gamma_{ab}$, $\gamma_{ac}$, $\gamma_{ad}$, $\Psi_c$, $\Psi_d$. 

\begin{figure}
    \centering
    \includegraphics[width=1\linewidth]{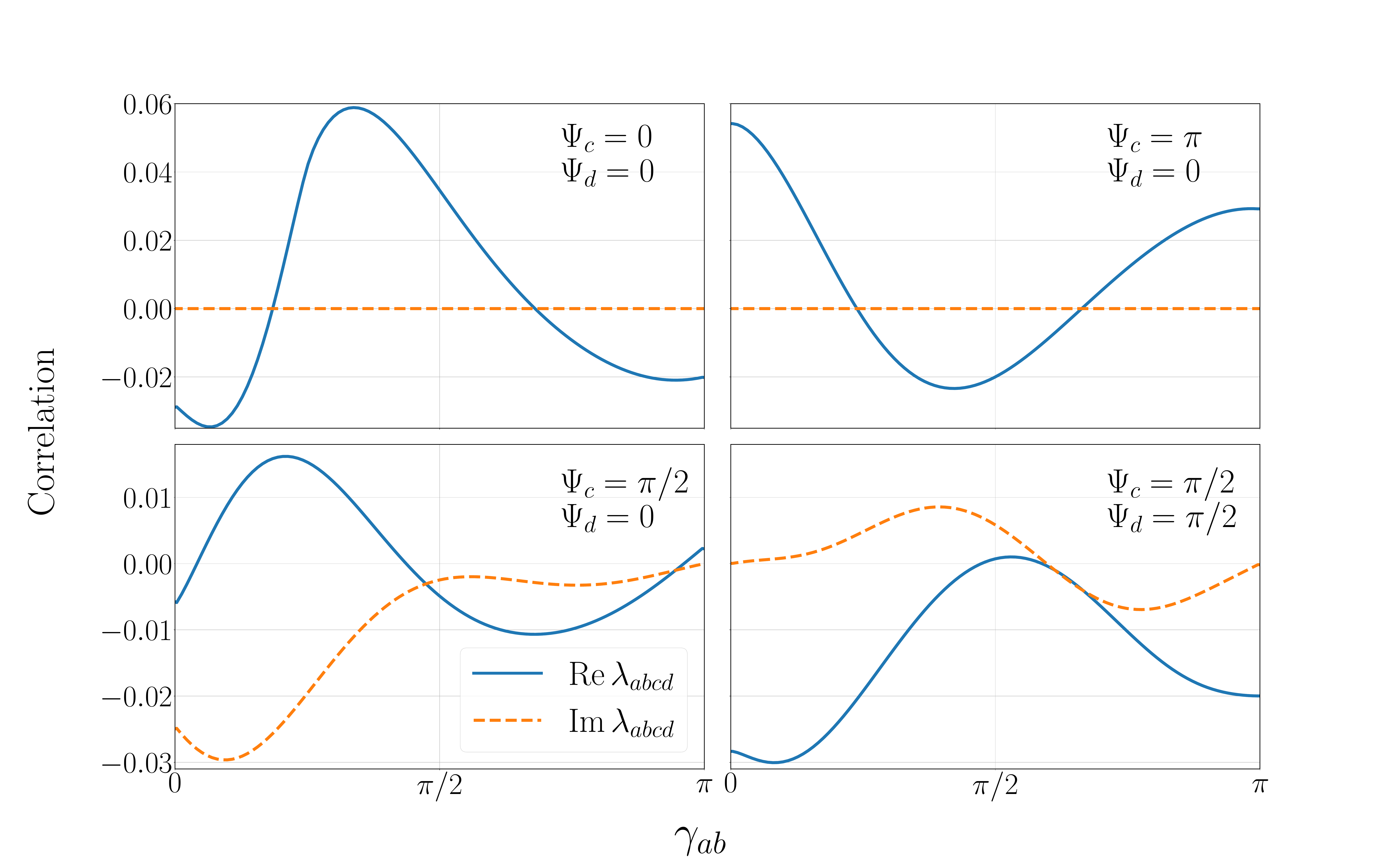}
    \caption{Real and imaginary parts of the function $\lambda_{abcd}$ as a function of $\gamma_{ab}$ for different values of the pulsars positions. We have fixed the values $\gamma_{ac} = \pi/4$, $\gamma_{ad} = 3 \pi/4$ and vary the phases $\Psi_c$, $\Psi_d$. }
    \label{fig:4PT}
\end{figure}

\begin{figure}
    \centering
    \includegraphics[width=1\linewidth]{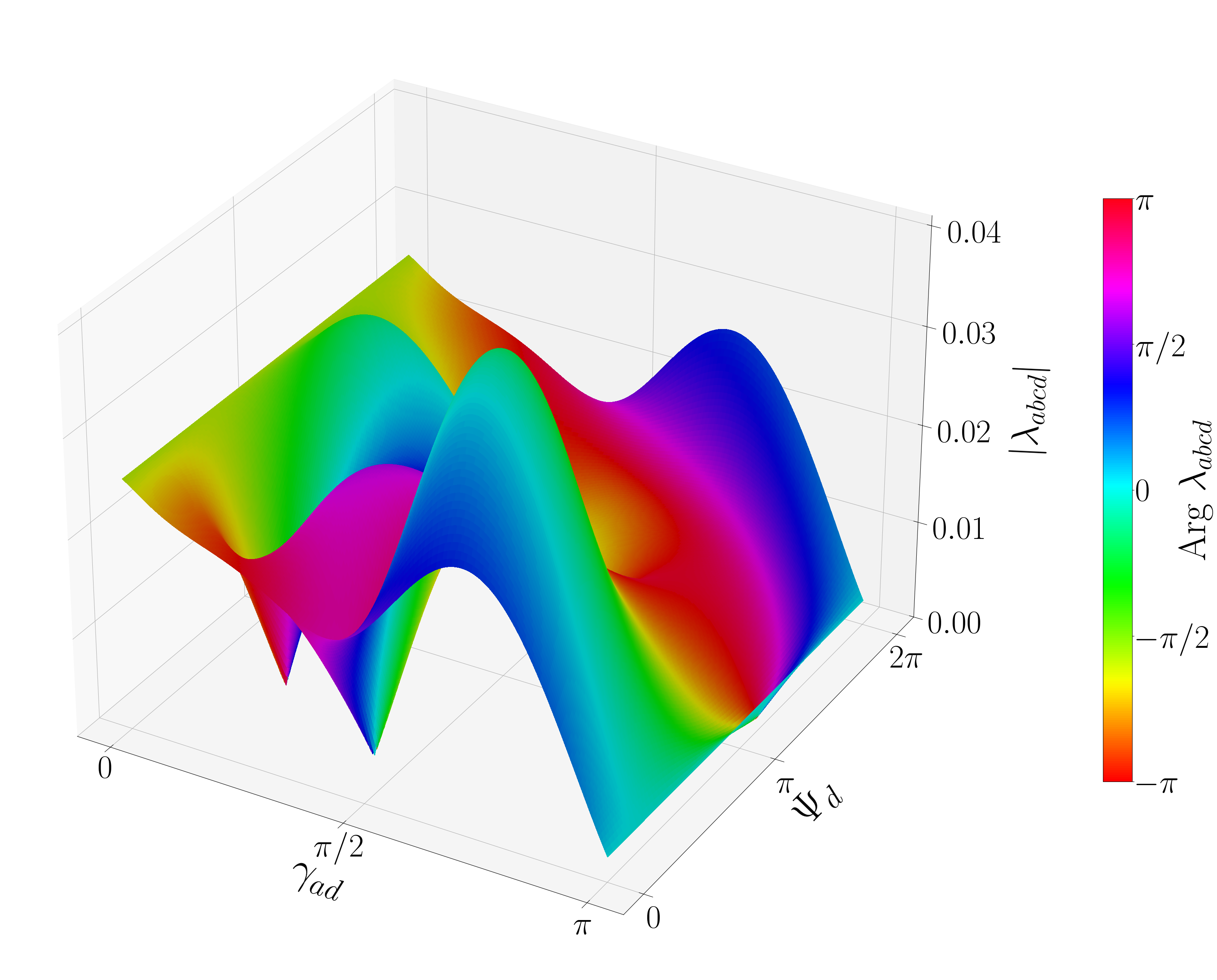}
    \caption{Modulus and phase of the function $\lambda_{abcd}$ as a function of $\gamma_{ad}$ and $\Psi_d$. We have fixed the values $\gamma_{ab} = \pi/4$, $\gamma_{ac} = 3 \pi/4$ and $\Psi_c = \pi/2$.}
    \label{fig:3dplot}
\end{figure}

The calculation performed in Appendix~\ref{app:lambdaabcd} shows that the 4PT correlator can be written as
\begin{align} \label{eq:lambdaabcd}
    \lambda_{abcd} &= G_0 + G_1 \log \bigg( \frac{1-\cos \gamma_{ab}}{2} \bigg) +  G_2 \log \bigg( \frac{1-\cos \gamma_{ad}}{2} \bigg) \nonumber \\
    &+  G_3 \log \bigg( \frac{1-\cos \gamma_{bc}}{2} \bigg) +  G_4 \log \bigg( \frac{1-\cos \gamma_{cd}}{2} \bigg),
\end{align}
where the angle $\gamma_{bc}$ between pulsars $b$ and $c$  can be obtained from the formula $\cos \gamma_{bc} = \cos \gamma_{ab} \cos \gamma_{ac} + \sin \gamma_{ab} \sin \gamma_{ac} \cos \Psi_c$ (and similarly for $\gamma_{cd}$). In Eq.~\eqref{eq:lambdaabcd}, the $G_i$'s are functions of the five pulsars angles, but they involve only $\cos$ and $\sin$ functions (possibly at the denominator).
For illustration purposes, we provide here the explicit expression of $G_1$:
\begin{widetext}
    \begin{equation}
    \label{eq:g1_expr}
        G_1=  \frac{e^{-\mathrm{i} \left(\Psi _c+\Psi _d\right)} \left(1-c_b\right) s_d^2 
   \left(s_c(1-c_b) - \left(1-c_c\right) s_b e^{\mathrm{i}  \Psi _c} \right) \left(2 s_b s_c e^{\mathrm{i}  \Psi
   _c}+\left(1+c_c\right) \left(1+c_b\right) e^{2 \mathrm{i}  \Psi _c}+\left(1-c_c\right)
   \left(1-c_b\right)\right)}{8 \left(1-c_c\right) \left(1+c_b\right) \left(s_d(1-c_b) - s_b \left(1-c_d\right) e^{\mathrm{i}  \Psi _d}\right)},
    \end{equation}
\end{widetext}
where $c_b= \cos \gamma_{ab}$, $s_b = \sin \gamma_{ab}$ and similarly for the other pulsars. 
The expressions of the other functions is too long to be displayed here (particularly $G_0$), but they are provided in the companion \textsc{Mathematica} notebook~\cite{mathematica}. On the other hand, the structure of the logarithmic terms in Eq.~\eqref{eq:lambdaabcd} is remarkably simple and involves only the pairwise angles between pulsars associated to a Fourier coefficient and a complex conjugate (see Eq.~\eqref{eq:deflambdaabcd}). Let us emphasize that our expressions for $\lambda_{abcd}$ are formally ill-defined for certain parameter values (see Appendix~\ref{app:lambdaabcd}), e.g. $\gamma_{ab}=0$. However, the \textit{limit} of $\lambda_{abcd}$ for $\gamma_{ab}\rightarrow0$ is well-defined and finite. We should not be surprised by such a behavior, since the HD correlation $\mu(\gamma)$ itself has this property when $\gamma \rightarrow 0$.
Looking at Eq.~\eqref{eq:g1_expr} it is worth noting that the only place where the complex number $\mathrm{i}$ appears in the 4PT function is through $e^{\mathrm{i} \Psi_c}$ and $e^{\mathrm{i} \Psi_d}$, so that $\lambda_{abcd}$ is real e.g. when $\Psi_c=\Psi_d=0$. 
In Figs.~\ref{fig:4PT} and~\ref{fig:3dplot}, we plot the shape of the 4PT function in some particular cases. 

The function  $\lambda_{abcd}$ obeys several properties under permutations of the pulsars, as expected from its definition in Eq.~\eqref{eq:deflambdaabcd}. These are:
\begin{equation}
\label{eq:lambda_symmetries}
\begin{aligned}
    \lambda_{abcd} &= \lambda_{cbad}, \\
    \lambda_{abcd} &= \lambda_{adcb}, \\
    \lambda_{abcd} &= (\lambda_{badc})^*.
\end{aligned}
\end{equation}
Notice that these symmetries also entail relations among the different $G_i$ functions, e.g. schematically $G_1(a \leftrightarrow c) = G_3$, $G_2(a \leftrightarrow c) = G_4$ and $G_1(b \leftrightarrow d) = G_2$.
We have verified that our computed 4PT function indeed satisfies these symmetries, which provides a non-trivial consistency check, given that the five angles $\gamma_{ab}$, $\gamma_{ac}$, $\gamma_{ad}$, $\Psi_c$, $\Psi_d$ transform in a complicated way under pulsar permutations.

To make these transformation properties transparent, it is convenient to 
reexpress the five angles appearing in the parametrization of 
Eqs.~\eqref{eq:coordinateSystem}—which was introduced in a specific 
reference frame—in a fully invariant form, namely in terms of scalar 
products of the physical direction vectors.

We begin with the polar angles. 
For each pulsar  $i=a,b,c,d$, we define
\begin{equation}
\label{eq:gamma_invariant}
\cos\gamma_{ai}
\equiv
\vec p_a \!\cdot\! \vec p_i ,
\end{equation}
so that $\gamma_{ai}$ corresponds directly to the angular separation 
between pulsars $a$ and $i$.

The azimuthal angle $\Psi_i$ admits a similarly invariant geometric interpretation. It is defined as the angle of the clockwise rotation around $\vec p_a$ required to bring pulsar $i$ into the plane singled out by $\vec p_a$ and $\vec p_b$, with positive projection along $\vec\delta_{ab} \equiv \vec p_b - \vec p_a$.

To make this definition explicit, we construct an orthonormal triad adapted to the physical configuration,
\begin{equation}
\label{eq:covariant_basis}
\mathbf e_z \equiv \vec p_a,
\qquad
\mathbf e_x \equiv
\frac{\vec p_b - (\vec p_b \!\cdot\! \vec p_a)\,\vec p_a}
{\sqrt{1-(\vec p_a \!\cdot\! \vec p_b)^2}},
\qquad
\mathbf e_y \equiv \mathbf e_z \times \mathbf e_x ,
\end{equation}
so that $\mathbf e_x$ lies in the $(\vec p_a,\vec p_b)$ plane and $\mathbf e_y$ completes a right–handed basis. The azimuthal phase is then obtained by projecting the component of $\vec p_i$ orthogonal to $\vec p_a$ onto the transverse plane. 
Introducing the complex (helicity) basis vector
\begin{equation}
\mathbf e_{+} \equiv \mathbf e_x + i\,\mathbf e_y ,
\end{equation}
one can write
\begin{equation}
\label{eq:phi_invariant}
e^{i\Psi_i}
\equiv
\frac{
\big(\vec p_i - (\vec p_a\!\cdot\!\vec p_i)\,\vec p_a\big)
\!\cdot\!
\mathbf e_{+}
}{
\sqrt{1-(\vec p_a\!\cdot\!\vec p_i)^2}
}.
\end{equation}

Substituting the coordinate expressions of Eq.~\eqref{eq:coordinateSystem} 
into Eqs.~\eqref{eq:gamma_invariant} and \eqref{eq:phi_invariant} 
one recovers precisely the angles introduced previously. The advantage of this formulation is that it is manifestly independent 
of any particular coordinate choice.
For convenience, we also provide in the \textsc{Mathematica} notebook~\cite{mathematica} a function computing the angles $\gamma_{ai}, \Psi_i$ from the unit vectors $\vec p_i$ using Eqs.~\eqref{eq:gamma_invariant}-\eqref{eq:phi_invariant}.

Under a permutation $\vec p_j \leftrightarrow \vec p_k$, the angles 
$\gamma_{ai}$ and phases $\Psi_i$ transform according to 
Eqs.~\eqref{eq:gamma_invariant} and \eqref{eq:phi_invariant}, once the 
basis vectors are written in the covariant form of 
Eq.~\eqref{eq:covariant_basis}. 
Evaluating different permutations of $\lambda_{abcd}$ allows us to explicitly verify the symmetry relations of Eqs.~\eqref{eq:lambda_symmetries}. 
Moreover, as explained in Appendix~\ref{sec:realfourier}, it also enables the computation of the four-point correlator in the real Fourier basis through expressions such as Eq.~\eqref{eq:aaaa}.

\subsubsection{Simplified expressions in particular cases}\label{subsubsec:simp}

When some of the four pulsars are equal to other ones, our expressions for the $G_i$ functions greatly simplify. Let us first examine the case $a=d$ ($a=b$ can then be trivially obtained by symmetry). We get 
\begin{align}
    \lambda_{abca} & =  \lim_{\gamma_{ad}\rightarrow 0} G_0 +   \lim_{\gamma_{ad}\rightarrow 0} G_3 \log \bigg( \frac{1-\cos \gamma_{bc}}{2} \bigg),
\end{align}
the limit of the other $G_i$ being zero. 
Next, in the case $a=c$, all the $G_i$'s are nonzero and contribute to the final result:
\begin{align}
    \lambda_{abca} & =  \lim_{\gamma_{ac}\rightarrow 0} G_0 +   \lim_{\gamma_{ac}\rightarrow 0} (G_1+G_3) \log \bigg( \frac{1-\cos \gamma_{ab}}{2} \bigg) \nonumber \\
    &+ \lim_{\gamma_{ac}\rightarrow 0} (G_2+G_4) \log \bigg( \frac{1-\cos \gamma_{ad}}{2} \bigg),
\end{align}
and we provide the corresponding simplified functions in the \textsc{Mathematica} file~\cite{mathematica}.
The most dramatic simplification happens when two pairs of pulsars are equal, so that there remains only one angle on which the final result depends. In the case where $a=d$ and $b=c$, we get
\begin{align}\label{eq:allen1}
    \lambda_{abba} = \frac{1}{120} \big( 13 + 10 \cos \gamma_{ab} + \cos^2 \gamma_{ab}  \big),
\end{align}
which matches the result obtain by Allen~\cite{Allen:2022dzg} in Eq.~F2, as it can be easily checked by noticing that the average in Eq.~F2 of Ref.~\cite{Allen:2022dzg} and Eq.~\eqref{eq:allen1} involve the same product of pattern functions. On the other hand, when $a=c$ and $b=d$, we get
\begin{align}
    \lambda_{abab} & = \frac{1}{120} \big( 113 + 10 \cos \gamma_{ab} -99 \cos^2 \gamma_{ab}  \big) \nonumber \\
    &+ \frac{1}{2} \big( 7 - 10 \cos \gamma_{ab} +3 \cos^2 \gamma_{ab} \big) \log \bigg( \frac{1-\cos \gamma_{ab}}{2} \bigg).
\end{align}
In the case where three pulsars are equal, we get
\begin{align}
    \lambda_{aaab} &= \frac{1}{240} \big( 75 + 23 \cos \gamma_{ab} - 45  \cos^2 \gamma_{ab} - 5  \cos^3 \gamma_{ab} \big) \nonumber \\
     &+ \frac{1}{2}(1-\cos \gamma_{ab}) \log \bigg( \frac{1-\cos \gamma_{ab}}{2} \bigg).
\end{align}
Finally, in the limit where all pulsars are equal, we get 
\begin{equation}\label{eq:auto}
     \lambda_{aaaa} = \frac{1}{5}.
\end{equation}
It is instructive to compare this last equation to the results in Ref.~\cite{Lamb:2025niq}, where terms of this form are computed. 
The setup considered in Ref.~\cite{Lamb:2025niq} is based on the same model outlined in Sec.~\ref{sec:bisp}. In particular, they compute the 4PT in the auto-pulsar case (exactly terms like Eq.~\eqref{eq:auto}), finding that the non-gaussianity contains contributions due to the finite observation time and the finite amount of sources. Moreover, they also consider the non-gaussianity arising from fluctuations in the total number of sources.\\
The finite observation time induces correlations among different frequencies~\cite{Crisostomi:2025vue}, which in Ref.~\cite{Lamb:2025niq} are modeled in terms of window functions.~\footnote{Notice that current PTA data analysis methods neglect these correlations. They are however partially mitigated by the timing model marginalization scheme (see Fig.~1 in Ref.~\cite{Crisostomi:2025vue}). See also~\cite{vanHaasteren:2014faa,Allen:2024uqs,Bernardo:2024tde, Bernardo:2025dat,Allen:2025waa}.} 
In our work, we are assuming for simplicity that different frequencies are uncorrelated. Therefore, we are not able to capture the non-gaussianity arising from the finite observation time correlating different frequencies, which is anyway expected to be subdominant as the typical form of the window function, namely $\omega(f_I - f_J) = \text{sinc}[\pi T (f_I - f_J)]$, is peaked around zero. \\
Thus, we can compare our results to Ref.~\cite{Lamb:2025niq} in two reference cases: the non-gaussianity arising from the finite amount of sources, with the total number of sources held fixed, and the non-gaussianity arising when fluctuations in the total number of sources are averaged over. Our description will be rather synthetic, and we refer to Ref.~\cite{Lamb:2025niq} for further details on their analysis.\\

In the former case, we can immediately use Eq.~\eqref{eq:4ptCorr}, as it does not involve any average over the total number of sources. By plugging Eq.~\eqref{eq:auto} and $\bar \mu (0) = 2/3$, we easily get:
\begin{equation}\label{eq:710}
    \frac{\langle \mathcal{C}_I^a  \mathcal{C}_{-I}^a  \mathcal{C}_I^a  \mathcal{C}_{-I}^a \rangle_{\phi,\vec\Omega}^\mathrm{NG}}{   \langle \mathcal{C}_I^a  \mathcal{C}_{-I}^a \rangle^2_{\phi, \vec\Omega}} = \frac{7}{10} \frac{\mathcal{H}_4^I}{(\mathcal{H}_2^I)^2},
\end{equation}
where the label $\text{NG}$ stands for the non-gaussian component. It is immediate to notice that Eq.~\eqref{eq:710} recovers the result obtained in Eq.~21 and Eq.~33 of Ref.~\cite{Lamb:2025niq}.\\

In the latter case, we need to implement the average over the total number of sources in Eq.~\eqref{eq:4ptCorr}. In order to do so, we follow the procedure in Ref.~\cite{Lamb:2025niq}, and group the total number of sources $N_I$ in a given frequency bin $I$ according to their amplitude. In formulae, we write 
\begin{equation}\label{eq:LambB}
    \mathcal{H}_2^I = \sum_B N_B \frac{(A_I^B)^2}{\omega_I^2} \qquad \mathcal{H}_4^I = \sum_B N_B \frac{(A_I^B)^4}{\omega_I^4},
\end{equation}
where each bin $B$ contains $N_B$ sources with typical amplitude $A_I^B$. Now, we use the fact that the number of sources within a given bin $B$ is Poisson-distributed, with mean $\bar N_B$~\cite{Sesana_2009, Sato:2024, Lamb:2025niq}.
Thus, the average over the total number of sources $N$ of the quantities in Eq.~\eqref{eq:LambB} will be
\begin{equation}\label{eq:avN}
    \la\mathcal{H}_2^I\ra_N = \sum_B \bar N_B \frac{(A_I^B)^2}{\omega_I^2} \qquad \la\mathcal{H}_4^I\ra_N = \sum_B \bar N_B \frac{(A_I^B)^4}{\omega_I^4}.
\end{equation}
Moreover, using the fact that the variance of a Poisson distributed variable is equal to the mean, we also get:
 \begin{equation}\label{eq:av2N}
    \la(\mathcal{H}_2^I)^2\ra_N = \la\mathcal{H}_2^I\ra_N^2 + \la\mathcal{H}_4^I\ra_N.
\end{equation}
We refer to Appendix~B of Ref.~\cite{Lamb:2025niq} for further details. 
Then, performing the average over the total number of sources and substituting Eqs.~\eqref{eq:avN} and \eqref{eq:av2N} into Eq.~\eqref{eq:4ptCorr}, it immediately follows that
\begin{align}\label{eq:4ptCorrN}
    &16 \langle \mathcal{C}_I^a  \mathcal{C}_{-J}^b  \mathcal{C}_K^c  \mathcal{C}_{-L}^d \rangle_{\phi, \vec\Omega,N}  = \delta_{IJ} \delta_{KL} \delta_{IK} ~\la\mathcal{H}_4^I\ra_N ~  \lambda_{abcd} +   \nonumber \\ 
    &+ \la\mathcal{H}_2^I\ra_N \la\mathcal{H}_2^K\ra_N  \bigg[ \delta_{IJ} \delta_{KL}  \bar \mu(\gamma_{ab}) \bar \mu(\gamma_{cd}) + \nonumber\\
    &+\delta_{IL} \delta_{JK} \bar \mu(\gamma_{ad}) \bar \mu(\gamma_{bc})    \bigg].
\end{align}
Considering the coincident limit and substituting Eq.~\eqref{eq:auto}, we then get
\begin{equation}\label{eq:2710}
    \frac{\langle \mathcal{C}_I^a  \mathcal{C}_{-I}^a  \mathcal{C}_I^a  \mathcal{C}_{-I}^a \rangle_{\phi, \vec\Omega, N}^\mathrm{NG}}{   \langle \mathcal{C}_I^a  \mathcal{C}_{-I}^a \rangle^2_{\phi, \vec\Omega, N}} = \frac{27}{10} \frac{\la\mathcal{H}_4^I\ra_N}{\la\mathcal{H}_2^I\ra_N^2},
\end{equation}
which recovers the result in Eq.~(20) of Ref.~\cite{Lamb:2025niq}.

The result in Eq.~\eqref{eq:4ptCorrN} naturally leads to interpreting $\lambda_{abcd}$ as the non-gaussianity arising after the average over the total number of sources. Depending on whether we are actually interested in performing the average over the total number of sources or not, we should then refer to Eqs.~\eqref{eq:4ptCorrN} or \eqref{eq:4ptCorr}, respectively.

\section{Connection to PTA data analysis}\label{sec:pta} %or marginalized likelihood
In Sec.~\ref{sec:4pt}, we computed the 4PT correlator of the Fourier coefficients describing the SGWB induced by a superposition of circularly inspiraling SMBHBs.\\
As we discussed, the 4PT correlator is the lowest order measure of non-gaussianity of an astrophysical SGWB, because the averaged 3PT correlator is vanishing.
In order to probe non-gaussianity in real data, we thus have to understand how the 4PT correlator actually enters the PTA parameter estimation pipeline. \\
To this purpose, we initially review the PTA pipeline, describing the sources of noise and how statistical inference on a gaussian SGWB is performed. Then, we extend the treatment to incorporate non-gaussianity. Technical results are presented in Appendices~\ref{sec:realfourier} and \ref{sec:marg}.

\subsection{Gaussian marginalized likelihood}\label{subsec:gauss}
Pulsar timing relies on detailed modeling of the propagation of radio pulses through a non-empty spacetime. Along their journey to Earth, the pulses are affected by several physical processes, including binary motion, interstellar and Solar System dispersion, Shapiro delays, and atmospheric effects~\cite{Edwards_2006, Shaifullah:2025pwz}. These contributions are encoded in a \textit{timing model} that relates the observed TOAs to the emission times at the pulsar, with its parameters estimated via a least-squares fit~\cite{Hobbs_2006}. 
Deviations between the observed TOAs and those predicted by the best-fit timing model define the \textit{timing residuals}. These residuals contain contributions from imperfectly determined timing-model parameters
 -- parameterized in terms of the \textit{design matrix} $\textbf{M}$, encoding the derivatives of the predicted TOAs with respect to the timing model parameters, i.e. introducing first order corrections to the best fit timing solution -- instrumental noise contributions, usually modeled as white noise,  as well as astrophysical processes. Examples include long-term stochastic variations in the pulsar spin frequency, which produce red-noise components, and the potential imprint of a SGWB.\\

For the sake of simplicity, we assume timing residuals only contain white noise, $\vec{n}$, and the contribution induced by the astrophysical SGWB previously described, $\vec{s}$. The reader is invited to refer to Refs.~\cite{Taylor:2021, chalumeau_phd, Shaifullah:2025pwz} for an extended discussion on the different contributions to the timing residuals, which however does not modify our conclusions. In formulae, 
\begin{equation}\label{eq:tr}
    \vec{\delta t}  = \vec{s} + \vec{n}.
\end{equation}
Assuming for simplicity that each of the $N_p$ pulsars has $N_\text{TOA}$ TOAs, the length of the vectors in the previous expression is $N_p \times N_\text{TOA}$. Following the previous indexing routine, we label by $x_{i}^{a}$ the $i$-th component of the quantity $\vec{x}$ for the pulsar $a$.\\
White noise is modeled as a zero-mean gaussian process, conventionally described by the covariance matrix~\cite{NANOGrav:2015qfw, Taylor:2021, Shaifullah:2025pwz} :
\begin{equation}\label{eq:white}
    \la n_{i, \mu}^a n_{j,\nu}^b\ra   =  \left[\left(E_\mu^2 \sigma_i^2 + Q_\mu^2\right)\delta_{ij}+ J_\mu^2 \delta_{e(i)e(j)}\right]\delta_{\mu\nu}\delta_{ab}.
\end{equation}
The left hand side stands for the correlation between the white noise timing delay at the $i$-th TOA of pulsar $a$, recorded by the receiver/backend system $\mu$, and that at the $j$-TOA of pulsar $b$, recorded by the receiver/backend system $\nu$. 
The quantities $E_\mu$, $Q_\mu$, and $J_\mu$ are commonly referred to as EFAC, EQUAD and ECORR, respectively, and are standard
timing noise parameters implemented in pulsar-timing software packages
such as \textsc{Tempo2}~\cite{Hobbs_2006, Edwards_2006} and \textsc{PINT}~\cite{Luo:2020ksx}.
The quantity $\sigma_i$ denotes the TOA uncertainty arising from the template-matching (radiometer-noise) fitting procedure. EFAC accounts for possible mis-calibration of the formal TOA uncertainties by rescaling the radiometer noise, while EQUAD represents an additional source of time-independent white
noise added in quadrature. These two contributions are assumed to be
uncorrelated between different TOAs. In contrast, ECORR models white noise that is fully correlated between the $i$-th and $j$-th
TOAs belonging to the same observing epoch, $e(i)=e(j)$, capturing effects
such as pulse-jitter noise.~\footnote{This can occur in multi-frequency observations where the total observing bandwidth is divided into several frequency subbands. In this case, a separate TOA is obtained from each subband, resulting in multiple TOAs at different radio frequencies associated with the same observing epoch.}

As the covariance is diagonal in the receiver/backend systems, we will omit the indices $\mu,\nu$ from now on. Therefore, we will simply label the covariance matrix element described by Eq.~\eqref{eq:white} as $\mathcal{N}^{ab}_{ij}$. 
While we show the explicit white-noise covariance
structure of Eq.~\eqref{eq:white} for illustration purposes, the following discussion applies to generic noise covariance matrices, provided they are
block diagonal in the pulsar indices $a, b$ as $\mathcal{N}^{ab}_{ij}$.~\footnote{If the PTA experiment does not time pulsars in different radio-frequency bands, then ECORR is completely degenerate with EQUAD, and the noise covariance matrix becomes exactly diagonal.} \\
Usually, also the SGWB is modeled as a zero-mean gaussian process. In particular, it is expanded in terms of a Fourier basis~\cite{Lentati:2014}: 
\begin{equation}\label{eq:sia}
\begin{split}
    \vec{s}^{a}_{i} &= \sum_{I=1}^{N_f}\left[a^a_I\cos\left(2\pi \frac{I}{T}t^a_i\right) + b_I^a\sin\left(2\pi \frac{I}{T}t^a_i\right) \right] \\
    &= \sum_{I=1}^{2N_f} F^a_{iI}~w_I^a,
\end{split}
\end{equation}
where $a_I^a, b_I^a$ are the real \textit{Fourier} coefficients, in contrast to the complex ones outlined in the previous sections, $T$ is the dataspan and $N_{f} $ labels the number of frequencies used in the expansion. In Eq.~\eqref{eq:sia}, we compactly packed the alternating basis of cosines and sines in the Fourier basis matrix $F^a$, and we defined $\vec{w}^a \equiv (a_1^a, b_1^a, a_2^a, b_2^a,~...)$. Notice that $\text{dim}(\vec{w}^a) = 2N_f$, as it contains both the $\vec{a}^a$ and the $\vec{b}^a$ coefficients. In the following, we will use both of the notations $\vec{w}$ and $\vec{a}, \vec{b}$ interchangeably.\\
In usual PTA data analysis, the Fourier coefficients are fully defined by their covariance matrix:~\footnote{As previously discussed, we neglect inter-frequency correlations.} 
\begin{equation}\label{eq:sgwbcov}
\begin{split}
    &\la a_I^a a_J^b\ra   = \la b_I^a b_J^b\ra   = \bar \mu(\gamma_{ab})\frac{S(f_I)}{T}\delta_{IJ}\\
    &\la a_I^a b_J^b\ra   = 0 ,
\end{split}
\end{equation}
where $S(f_I)$ is the one-sided power spectral density of SGWB, which reads
\begin{equation}\label{eq:sf}
    S(f)  \equiv \frac{h_c^2(f)}{12\pi^2 f^3} \equiv \frac{A^2_\text{gw}}{12\pi^2 f_0^3}\left(\frac{f}{f_0}\right)^{-\gamma_\text{gw}}.
\end{equation}
This relates the statistics of the Fourier coefficients to the power spectral density of the SGWB (or, equivalently, to its \textit{characteristic strain} $h_c(f)$), which is commonly adopted in PTA studies.
Here, $A_\text{gw}$ is the amplitude of the power spectral density, normalized at the pivotal frequency $f_0$, which is typically chosen to be either $\text{(1 yr)}^{-1}$ or $\text{(10 yr)}^{-1}$. The spectral index is labeled by $\gamma_\text{gw}$, and it is $13/3$ for the SGWB induced by circularly inspiraling SMBHBs. In the following, we will conveniently define the SGWB covariance matrix as
\begin{equation}\label{eq:sgwbcov2}
    \Phi^{ab}_{IJ} \equiv \la w_I^a~w_J^b\ra   ,
\end{equation}
whose matrix elements are set by Eq.~\eqref{eq:sgwbcov}. \\

The likelihood for the timing residuals in Eq.~\eqref{eq:tr} can thus be written as:
\begin{equation}\label{eq:lik}
    \mathcal{L}(\vec{\delta t}|\vec{w}, \vec{\eta}) = \frac{\exp\left[-\frac{1}{2}\left(\vec{\delta t} - F\vec{w}\right)^T \mathcal{N}^{-1} \left(\vec{\delta t} - F\vec{w}\right)\right]}{\sqrt{(2\pi)^N|\mathcal{N}|}},
\end{equation}
in terms of the vector of Fourier coefficients $\vec{w}$ and of the vector of hyperparameters $\vec{\eta}$, containing e.g. the white noise parameters in Eq.~\eqref{eq:white} and the amplitude and slope of the power spectral density in Eq.~\eqref{eq:sf}. Notice that all the quantities in Eq.~\eqref{eq:lik} are referred to the entire array of pulsars; namely, $\text{dim} (\vec{\delta t}) \equiv N= N_p \times N_\text{TOA}$, $\text{dim}(\vec{w})\equiv W = N_p\times 2N_f$ and $F, \mathcal{N}$ are block-diagonal matrices with dimensions $\text{dim}( F) = (N, W)$ and $\text{dim}(\mathcal{N}) = (N, N)$, respectively. In order to perform parameter estimation, we need a prior on the Fourier coefficients and on the hyperparameters. Schematically, in fact, the Bayes theorem~\cite{Bayes:1763} ensures that the posterior $\mathcal{P}$ is linked to the likelihood $\mathcal{L}$ via
\begin{equation}
    \mathcal{P}(\vec{w}, \vec{\eta}| \vec{\delta}t  ) \propto \mathcal{L}(\vec{\delta}t | \vec{w}, \vec{\eta})\Pi_\text{G}(\vec{w}|\vec{\eta})\Pi(\vec{\eta}),
\end{equation}
where the $\Pi_\text{G} ,~\Pi$ terms represent priors. In the following, we will be concerned with the term $\Pi_\text{G}(\vec{w}|\vec{\eta})$, which describes the prior on the Fourier coefficients $\vec{w}$ given the hyperparameters $\vec{\eta}$ of the analysis. We recall that, in the gaussian approximation, the SGWB is fully described by the covariance matrix~\eqref{eq:sgwbcov2}. Therefore, we can write:
\begin{equation}\label{eq:prior}
    \Pi_\text{G}(\vec{w}|\vec{\eta}) = \frac{\exp (-\frac{1}{2}\vec{w}^T \Phi^{-1} \vec{w})}{\sqrt{(2\pi)^W|\Phi|}},
\end{equation}
where $\Phi$ is the covariance matrix described by Eqs.~\eqref{eq:sgwbcov}-\eqref{eq:sgwbcov2}.
However, in PTAs, we are typically not interested in the specific value of the Fourier coefficients, as they describe a single realization of the stochastic process. Rather, we are interested in the hyperparameters $\vec{\eta}$ defining the distribution which the Fourier coefficients $\vec{w}$ are sampled from. In probability theory, this statement corresponds to marginalizing over the $\vec{w}$ parameters, to obtain a posterior of just the $\vec{\eta}$ quantities. Therefore, we end up with 
\begin{equation}\label{eq:marglik}
    \begin{split}
      \mathcal{P}(\vec{\eta}| \vec{\delta}t  )  = &\int d\vec{w}~ \mathcal{P}(\vec{w}, \vec{\eta}| \vec{\delta}t  ) \propto \\
      &\int d\vec{w}~ \mathcal{L}(\vec{\delta}t | \vec{w}, \vec{\eta})\Pi_\text{G}(\vec{w}|\vec{\eta})\equiv \mathcal{L}^\text{G}_\text{marg}(\vec{\delta t}|\vec{\eta}),
      \end{split}
\end{equation}
where the gaussian \textit{marginalized likelihood} $\mathcal{L}^\text{G}_\text{marg}(\vec{\delta t}|\vec{\eta})$ is only conditioned on the hyperparameters $\vec{\eta}$. Using Eqs.~\eqref{eq:lik} and \eqref{eq:prior}, then, we find 
\begin{equation}
\label{eq:marg_gauss_like}
    \mathcal{L}^\text{G}_\text{marg}(\vec{\delta t}|\vec{\eta}) \propto \exp\left[-\frac{1}{2}\vec{\delta t}^T\left(\mathcal{N} + F \Phi F^T\right)^{-1}\vec{\delta t}\right].
\end{equation}
We present a technical derivation of this relation in Appendix~\ref{subsec:agauss}, along the lines of Appendix B in Ref.~\cite{Falxa:2025}.

\subsection{Non-gaussian marginalized likelihood}\label{subsec:nglik}
We now consider the case of a non-gaussian SGWB. 
In practice, accounting for non-gaussianity means that the gaussian prior in Eq.~\eqref{eq:prior} will not encode the full statistics of the Fourier coefficients describing the SGWB. Therefore, we need to modify the prior in Eq.~\eqref{eq:prior} to account for higher order contributions. A convenient way to do that is by using the multivariate Edgeworth expansion~\cite{Kitaura_2012}, which reads:
\begin{widetext}
\begin{equation}\label{eq:ngprior}
%\begin{split}
    \Pi(\vec w|\vec{\eta}) \propto \Pi_\text{G}(\vec w|\vec \eta) \times\left[1 + \frac{1}{4!}\sum_{\{a'\}}^{N_p}\sum_{\{I'\}}^{2 N_f}\sum_{\{a\}}^{N_p}\sum_{\{I\}}^{2 N_f}  (\kappa_4)^{a'b'c'd'}_{I'J'K'L'} (\Phi^{-1/2})^{aa'}_{II'}(\Phi^{-1/2})^{bb'}_{JJ'}(\Phi^{-1/2})^{cc'}_{KK'}(\Phi^{-1/2})^{dd'}_{LL'} H^{abcd}_{IJKL}(\vec{\nu}) \right],    
    %\Pi(\vec w|\vec{\eta}) \propto& \Pi_\text{G}(\vec w|\vec \eta)\times \left[1 + \frac{1}{4!}\sum_{a'...}\sum_{i'..}(\kappa_4)^{a'...}_{i'...}\right.\\
    %\times&\left.\sum_{a...}\sum_{i...}(\Phi^{-1/2})^{aa'}_{ii'}...H^{a ...}_{i ...}(\vec{w})\right]
%\end{split}
\end{equation}
\end{widetext}
up to fourth order, where $H_{IJKL}^{abcd}$ are the fourth-order Hermite polynomials, defined in terms of the unit-variance variable $\vec\nu$:
\begin{equation}\label{eq:nuI}
    \nu^a_I = \left(\Phi^{-1/2} \vec{w} \right)^{a}_{I}
\end{equation}
and we conveniently group the indices as $\{a'\} \equiv\{a',b',c',d'\}$, for instance. In Eq.~\eqref{eq:ngprior}, we explicitly highlight that the indices $\{a\}, \{a'\}$ run over the total number of pulsars $N_p$ in the considered dataset, whereas the indices $\{i\}, \{i'\}$ run up to $2N_f$. For simplicity, we omit this explicit specification in the following. 
We present detailed expressions and extended discussions in Appendix~\ref{sec:ang}. The Edgeworth series is an asymptotic expansion, and displays pathological behaviors in the tails of the distribution, leading e.g. to negative probability densities. Therefore, we will assume that the level of non-gaussianity is relatively small compared to the gaussian background. This assumption is reasonable for the lowest frequency bins, as the expected number of SMBHBs contributing to the background is rather large. However, it may fail at larger frequencies, where in any case even the usual power law approximation of the spectrum fails~\cite{Sesana_2008}. \\
In fact, at the upper end of PTA frequencies, $\mathcal{O}(1)$ sources are expected to dominate the signal, giving rise to a strongly non-gaussian regime. In this case, from a data analysis perspective, it becomes more appropriate to model the signal as deterministic Continuous Gravitational Waves (CGWs) on top of the SGWB (e.g. see Ref.~\cite{IPTA:2023ero}).~\footnote{We thank Mikel Falxa for pointing this out.}   
On the other hand, the large number of contributing SMBHBs in the lowest frequency bins implies that any non-gaussian features are likely to be strongly suppressed and thus difficult to detect.\\
We therefore expect our analysis to be most relevant in the intermediate frequency regime (roughly around $10^{-8}~\text{Hz}$), where the number of contributing SMBHBs remains substantial but not overwhelmingly large. For an illustrative example, see Fig.~5 of Ref.~\cite{Lamb:2025niq}, which quantifies the expected level of non-gaussianity in the pulsar autocorrelation across different frequency bins.
\\
The quantity $(\kappa_4)^{a'...}_{I'...}$ is the \textit{connected} 4PT correlator of Fourier coefficients; namely,
\begin{equation}\label{eq:kappa4}
\begin{split}
    &(\kappa_4)^{abcd}_{IJKL} \equiv \la w^{a}_{I}w^{b}_{J}w^{c}_{K}w^{d}_{L}\ra  _c  = \\
    &=\la w^{a}_{I}w^{b}_{J}w^{c}_{K}w^{d}_{L}\ra   - \la w^{a}_{I}w^{b}_{J}\ra  \la w^{c}_{K}w^{d}_{L}\ra  -\\
    &\,\,-\la w^{a}_{I}w^{c}_{K}\ra  \la w^{b}_{J}w^{d}_{L}\ra  - \la w^{a}_{I}w^{d}_{L}\ra  \la w^{b}_{J}w^{c}_{K}\ra  .
\end{split}
\end{equation}
We may be tempted to immediately relate this quantity to the 4PT function computed in Sec.~\ref{sec:4pt}, see Eq.~\eqref{eq:4ptCorr}. However, while the 4PT of Sec.~\ref{sec:4pt} is expressed in terms of the \textit{complex} Fourier coefficients, here we are dealing with \textit{real} Fourier coefficients. The relation between the 4PT in the two different basis is worked out in Appendix~\ref{sec:realfourier}.\\ 
As the relation between the two basis can be easily figured out, importantly, the results obtained in Sec.~\ref{sec:4pt} allow us to directly compute the connected 4PT. Therefore, substituting the findings of Sec.~\ref{sec:4pt} in Eqs.~\eqref{eq:ngprior}-\eqref{eq:kappa4}, we obtain the most general form of the non-gaussian prior, up to fourth order in the Fourier coefficients. \\
From this point on, the discussion proceeds exactly as in Sec.~\ref{subsec:gauss}. Namely, we can compute a non-gaussian marginalized likelihood, which schematically reads

\begin{widetext}
\begin{equation}\label{eq:ngmarg}
    \begin{split}
    &\mathcal{L}_\text{marg}(\vec{\delta t}|\vec{\eta}) = \mathcal{L}^\text{G}_\text{marg}(\vec{\delta t}|\vec{\eta}) \left\{1 + \frac{1}{4!}\sum_{\{a\}}\sum_{\{I\}} (\kappa_4)^{abcd}_{IJKL}\left[(\Phi^{-1}\hat w)^{a}_{I}(\Phi^{-1}\hat{w})^{b}_{J}(\Phi^{-1}\hat{w})^{c}_{K}(\Phi^{-1}\hat{w})^{d}_{L} \right.\right.\\
    &+\left.\left.~6~\left((\Phi^{-1}\hat w)_{I}^{a}(\Phi^{-1}\hat w)_{J}^{b} \left(\Phi^{-1}\Sigma^{-1}\Phi^{-1}\right)_{KL}^{cd}  -\left(\Phi^{-1}\Sigma^{-1}\Phi^{-1}\right)_{IJ}^{ab}(\Phi^{-1})^{cd}_{KL} - (\Phi^{-1}\hat w)_{I}^{a} (\Phi^{-1}\hat w)_{J}^{b}(\Phi^{-1})^{cd}_{KL}\right)  \right.\right.\\
    &+\left.\left.~3~\left(\left(\Phi^{-1}\Sigma^{-1}\Phi^{-1}\right)_{IJ}^{a b}\left(\Phi^{-1}\Sigma^{-1}\Phi^{-1}\right)_{KL}^{c d} + (\Phi^{-1})^{ab}_{IJ}(\Phi^{-1})^{cd}_{KL}\right)\right]\right\},
    \end{split}
\end{equation}
\end{widetext}
with 
\begin{equation}
\begin{split}\label{eq:sub}
    &\Sigma = F^T\mathcal{N}^{-1}F + \Phi^{-1}\\
    &\hat w = \Sigma^{-1}F^T \mathcal{N}^{-1}\vec{\delta t}.
\end{split}
\end{equation}
Full expressions and derivations, alongside with extended discussions,  are presented in Appendix~\ref{sec:ang}. Assuming different frequencies to be uncorrelated, we show in Appendix~\ref{sec:ang} that the total number of non-zero terms entering the sum in Eq.~\eqref{eq:ngmarg} is of the order of $\mathcal{O}(N_p^4 \times N_f)$. This quantity grows rapidly with the number of pulsars and can substantially increase the computational cost. Although optimizing this aspect is beyond the scope of this work, we outline a possible workaround below Eq.~\eqref{eq:intre} and comment further on this aspect in Sec.~\ref{sec:concl}.
To sum up, the marginalized likelihood in Eq.~\eqref{eq:ngmarg}, complemented with the definitions in Eq.~\eqref{eq:sub} and Appendix~\ref{sec:realfourier}, formally demonstrates how non-gaussianity can be inserted, at lowest order, in the PTA data analysis pipeline.

\section{Conclusions}~\label{sec:concl}
A common assumption in PTA searches is that the SGWB is gaussian, meaning that it can be fully characterized by its 2PT correlation function. While this assumption may be well justified for smooth backgrounds, it can break down in the scenario considered in the present work, namely an astrophysical SGWB produced by the incoherent superposition of a large number of SMBHBs. In this case, it becomes essential to investigate higher-point correlators.   \\

In this work, we have derived the complete 4PT correlator of the Fourier coefficients describing an astrophysical SGWB as a function of the angular position of four pulsars, capturing the leading-order non-gaussian contribution. As shown in Sec.~\ref{sec:4pt}, the result naturally separates into two parts: a gaussian component, proportional to the square of the 2PT correlator, and a genuinely non-gaussian term, namely the \textit{connected} 4PT correlator.\\

This quantity is central for searches of non-gaussianity in real PTA data. In the same way the shape of the HD correlation is crucial for distinguishing a SGWB from a common red noise process, any search for non-gaussian features requires the four-pulsar analogue of the HD curve. We denote its non-trivial component by $\lambda_{abcd}$. This function is introduced in Sec.~\ref{sec:4pt}, where we also discuss its main properties, while its detailed derivation is presented in Appendix~\ref{app:lambdaabcd}. The full expression of $\lambda_{abcd}$ appears in Eq.~\eqref{eq:lambdaabcd}. The associated functions $G_i$ are particularly lengthy and are thus presented in a companion \textsc{Mathematica} notebook~\cite{mathematica}.

To validate our calculation, we perform several consistency checks. First, we have checked our explicit formula against direct numerical integrations, finding a perfect agreement. Second, we verified that the 4PT correlator satisfies symmetry properties under specific permutations of the pulsars, which is a non-trivial check as the angular variables transform intricately under such exchanges. Then, we checked that our findings reproduce expressions presented in Refs.~\cite{Allen:2022dzg} and~\cite{Lamb:2025niq} in the appropriate limits (see Sec.~\ref{subsubsec:simp}). Notably, comparison with Ref.~\cite{Lamb:2025niq} reveals that $\lambda_{abcd}$ can straightforwardly be interpreted as the non-gaussianity arising after averaging over fluctuations in the total number of sources characterizing the SMBHB population. If this average is not performed, the expression of the non-gaussian contribution also contains terms proportional to the HD function, as shown in Eq.~\eqref{eq:4ptCorr}.\\

Although the angular dependence of the 4PT correlator is derived within the simplified toy model introduced in Sec.~\ref{sec:bisp}, we argue that the result has broader validity. In fact, analogously to the HD, $\lambda_{abcd}$ arises from angular averages of products of antenna pattern functions (see Appendix~\ref{app:lambdaabcd} for further details). Consequently, its form is expected to be insensitive to the specific physical mechanism generating the background, provided it consists of GWs.\\

Then, in Sec.~\ref{sec:4pt}, we show how the 4PT correlator enters the PTA parameter estimation pipeline. To this end, we derive in Eq.~\eqref{eq:ngmarg} a marginalized likelihood that perturbatively incorporates the contribution of the 4PT correlation function, with the detailed derivation presented in Appendices~\ref{sec:realfourier} and~\ref{sec:marg}.

Because our approach relies on a perturbative expansion, it is not expect to be reliable at the highest PTA frequencies, where the signal is generated by $\mathcal{O}(1)$ sources and is therefore strongly non-gaussian~\cite{Sesana_2008}. Conversely, while our results are valid for the lowest PTA frequencies, the large number of SMBHBs effectively suppresses the amount of observable non-gaussianity~\cite{Lamb:2025niq}. Therefore, we anticipate that our results will be most relevant in the intermediate frequency regime (roughly around $10^{-8}~\text{Hz}$), where the number of contributing SMBHBs is significant but not excessively large. 

Although the derivation is mathematically consistent and properly accounts for the 4PT contribution, the practical evaluation of the marginalized likelihood in Eq.~\eqref{eq:ngmarg} may be computationally demanding. In Appendix~\ref{sec:ang}, we discuss possible strategies that might help alleviating the computational burden. We also identify a set of dimensionless parameters $\epsilon_I$ as quantities to be inferred from data, quantifying the strength of the non-gaussian contribution relative to the (square of) the gaussian one within the $I$-th frequency bin. Anyway, we leave the question of the optimal strategy for a cost-effective implementation of Eq.~\eqref{eq:ngmarg}  in the PTA pipeline for future investigation.
In passing, we remark that a particularly interesting avenue could be to identify which pulsar configurations dominate the sum appearing in Eq.~\eqref{eq:ngmarg} and restrict the analysis to those subsets. Along with the broader implementation challenges, this issue is also deferred to future investigation.\\

Finally, although we argued that the angular structure of the 4PT should be robust beyond the specific toy model considered here, the overall amplitude of the non-gaussian component (for instance $\epsilon_I$ or, more generally $H_4^I$ in Eq.~\eqref{eq:intre}) depends on the properties of the source generating the background. It would therefore be interesting to explore more realistic SMBHB population models than the one considered in Sec.~\ref{sec:bisp}, for instance by including orbital eccentricity or allowing for non-face-on binaries. We leave this extension for future work.

\section*{Acknowledgements}
We warmly thank Mikel Falxa, William Lamb and Cecilia Sgalletta for their useful comments and discussions on the manuscript.
A.K. thanks the Fundação para a Ciência e Tecnologia (FCT), Portugal, for the financial support to the Center for Astrophysics and Gravitation (CENTRA/IST/ULisboa) through grant No. UID/PRR/00099/2025 and grant No. UID/00099/2025, as well as to the FCT project ``Gravitational waves as a new probe of fundamental physics and astrophysics'' grant agreement 2023.07357.CEECIND/CP2830/CT0003. The work of C.S. was supported by the European Union – NextGeneration EU, mission 4, component 1, CUP C93C24004930006. M.V. acknowledges support from the PRIN 2022 Grant “GUVIRP
- Gravity tests in the UltraViolet and InfraRed with Pulsar timing” and the European Union’s Horizon ERC Synergy Grant “Making Sense of the Un-
expected in the Gravitational-Wave Sky” (Grant No.
GWSky-101167314).

\newpage
\onecolumngrid
\appendix

\section{Calculation of $\lambda_{abcd}$} \label{app:lambdaabcd}
Using the definition in Eq.~\eqref{eq:deflambdaabcd} and the expression of $c_{I,i}^a$ in Eq.~\eqref{eq:defcI}, we find that the 4PT function can be expressed as the average over the source direction $\vec \Omega$:
\begin{equation}\label{eq:lambdaapp}
    \lambda_{abcd} = \bigg \langle \big( F_a^+(\vec \Omega) - \mathrm{i}  F_a^\times(\vec \Omega) \big) \big( F_b^+(\vec \Omega) + \mathrm{i}  F_b^\times(\vec \Omega) \big) \big( F_c^+(\vec \Omega) - \mathrm{i}  F_c^\times(\vec \Omega) \big) \big( F_d^+(\vec \Omega) + \mathrm{i}  F_d^\times(\vec \Omega) \big) \bigg \rangle_{\vec \Omega},
\end{equation}
where the angular pattern functions $F_a^A$ are defined in Eq.~\eqref{eq:defFa}. Notice that this result is independent on the choice of the polarization angle $\psi$, which modifies the pattern functions as $F_a^+ - \mathrm{i} F_a^\times \rightarrow e^{-2 \mathrm{i} \psi} (F_a^+ - \mathrm{i} F_a^\times)$. 
In the particular coordinate system defined in Eq.~\eqref{eq:coordinateSystem}, we express the SMBHB source direction as
\begin{align}
    \vec \Omega &= \begin{pmatrix}
        \sin \theta \cos \beta \\ \sin \theta \sin \beta  \\  \cos \theta
    \end{pmatrix} \; , \quad  
    \vec m = \begin{pmatrix}
        \sin \beta \\- \cos \beta  \\  0
    \end{pmatrix} \; , \quad 
    \vec n = \begin{pmatrix}
        \cos \theta \cos \beta \\ \cos \theta \sin \beta  \\  -\sin \theta
    \end{pmatrix} \; .
\end{align}
In order to compute $\lambda_{abcd}$, we have thus to integrate over the two angles $\theta$, $\beta$. Plugging in the different expressions, Eq.~\eqref{eq:lambdaapp} becomes:
\begin{align} \label{eq:defLambda}
    \lambda_{abcd} &= \frac{1}{4\pi} \int_{0}^\pi \sin \theta \mathrm{d} \theta
 \int_0^{2\pi} \mathrm{d}\beta \, \sin^2 \frac{\theta}{2} \frac{f_b(\theta, \beta)^2 (f_c(\theta, \beta)^2)^* f_d(\theta, \beta)^2 }{8 g_b(\theta, \beta) g_c(\theta, \beta) g_d(\theta, \beta)},
 \end{align}
 where the functions at the numerator and the denominator read e.g. for pulsar $d$:
 \begin{align}
     f_d(\theta, \beta) &= \cos \gamma_{ad} \sin \theta -  \sin \gamma_{ad} \cos \theta \cos (\beta - \Psi_d )   + \mathrm{i} \sin \gamma_{ad} \sin(\beta - \Psi_d)  \\
     g_d(\theta, \beta) &= 1 + \cos \gamma_{ad} \cos \theta + \sin \gamma_{ad} \sin \theta \cos (\beta - \Psi_d). \label{eq:numeratorLambda}
 \end{align}
The other functions $f_b, f_c, g_b, g_c$ take the same form, with the replacement $\Psi_d \rightarrow 0$ for pulsar $b$. We will now analytically integrate this expression, using first the residue theorem to compute the integral in $\beta$ and then simple integration formulae to deal with the integral in $\theta$. 

Let us first begin with the integral in $\beta$. We use the change of variable $z = e^{\mathrm i \beta}$ ($\mathrm{d} \beta = -i \mathrm{d}z/z$) in order to transform it to an integral along the unit circle in the complex plane, which allows us to use the residue theorem. There remains to find the poles of the integrand inside the unit circle. To this aim, let us first rewrite the denominator in order to find its zeros:
 \begin{align}
     g_d(\theta, \beta) &= \frac{e^{-\mathrm{i} \Psi_d} \sin \gamma_{ad} \sin \theta}{2 z} \big( z-z_d \big) \bigg(z-\frac{1}{z_d^*} \bigg) \; , \\
     z_d &= - e^{\mathrm{i} \Psi_d} \frac{1 + \cos \theta}{\sin \theta} \frac{1 + \cos \gamma_{ad}}{\sin \gamma_{ad}} \label{eq:zd}
 \end{align}
 and similarly for the other denominators. We also rewrite the numerators as
 \begin{align}
     f_d(\theta, \beta) &= \frac{e^{-\mathrm{i} \Psi_d} \sin \gamma_{ad} (1-\cos \theta)}{2z} \big( z - z_d \big) \big( z - y_d \big) \; , \\
     y_d &=  e^{\mathrm{i} \Psi_d} \frac{ 1+\cos \theta }{ \sin \theta }  \frac{1- \cos \gamma_{ad} }{\sin \gamma_{ad}}.
 \end{align}
 Notice the remarkable property that one zero of the numerator coincides with a zero of the denominator. This holds for pulsar $c$ as well, since
 \begin{equation}
     f_c^*(\theta, \beta) = - \frac{e^{-\mathrm{i} \Psi_c} \sin \gamma_{ac} (1+\cos \theta)}{2z} \bigg( z - \frac{1}{z_c^*} \bigg) \bigg( z - \frac{1}{y_c^*} \bigg).
 \end{equation}
 Thus, for each denominator $g_b$, $g_c$, $g_d$, only one of the two zeros survives and gives a contribution to the residue theorem. All in all, the $z$ integral is given by the sum over residues at three simple poles -- $1/z_b^*$, $1/z_d^*$ and $z_c$ -- plus a pole of order 4 at $z=0$. All of the residues are readily computed using \textsc{Mathematica}. 

Although it has the longest expression, the residue in $z=0$ is remarkably simple, as it reduces to a simple polynomial in $\cos \theta$. The outer integral in $\theta$ is thus trivial to compute for this term. The final form of this residue is provided in the companion notebook.

 The residues at the simple poles are more tricky to deal with. In particular, depending on the parameter values, the poles do not always lie inside the unit circle, and therefore the contour integral over $z$ does not always include their contributions. Looking at the expression~\eqref{eq:zd}, it turns out that:
 \begin{itemize}
     \item $1/z_b^*$ is inside the unit circle for $0 \leq \theta < \pi - \gamma_b$;
     \item $z_c$ is inside the unit circle for $\pi - \gamma_c < \theta \leq \pi$;
     \item $1/z_d^*$ is inside the unit circle for $0 \leq \theta < \pi - \gamma_d$.
 \end{itemize}
We can thus split the integral defining $\lambda_{abcd}$ in~\eqref{eq:defLambda} into three separate pieces, each associated with one of the three poles. For each term, the integration over $\theta$ is restricted to the range in which the corresponding pole lies inside the unit circle and thus yields a non-vanishing contribution to the $z$ integral. Explicitly, we write
\begin{equation} \label{eq:residuesSimplePoles}
    \lambda_{abcd} = 2\pi \mathrm i  \mathrm{Res}_0 + \int_{-c_b}^1 \mathrm{d}x B(x) + \int_{-1}^{-c_c} \mathrm{d}x C(x) + \int_{-c_d}^1 \mathrm{d}x D(x) 
\end{equation}
where $x=\cos \theta$, $c_b = \cos \gamma_{ab}$, $c_c = \cos \gamma_{ac}$, $c_d=\cos \gamma_{ad}$, $\mathrm{Res}_0$ stands for the residue in $z=0$ integrated in $x$ and the three residues at simple poles are given by 
\begin{align}
    B(x) &= \frac{(c_b+x) e^{-\mathrm{i} (\Psi _c+\Psi _d)}}{32 (1+x) \left(1-c_b^2\right) \Gamma_1(x) \bigg( \sqrt{\left(1+c_b\right)
   \left(1-c_d\right)} e^{\mathrm{i} \Psi _d} -\sqrt{\left(1-c_b\right) \left(1+c_d\right)} \bigg) } \\
    &\times \left(\sqrt{\left(1-c_c\right) \left(1+c_b\right)} e^{\mathrm{i} \Psi  _c}-\sqrt{\left(1+c_c\right) \left(1-c_b\right)}\right) \left(\sqrt{\left(1+c_c\right) \left(1+c_b\right)} e^{\mathrm{i}
   \Psi _c}+\sqrt{\left(1-c_c\right) \left(1-c_b\right)}\right)^2 \nonumber \\
   &\times \left((1+x)  \sqrt{\left(1+c_b\right)
   \left(1-c_d\right)} e^{\mathrm{i} \Psi _d} + (1-x)
   \sqrt{\left(1-c_b\right) \left(1+c_d\right)}\right)^2 \nonumber \\
   &\times \bigg(  (1+x)
   \sqrt{\left(1+c_b\right) \left(1+c_d\right)} e^{\mathrm{i} \Psi _d} - (1-x) \sqrt{\left(1-c_b\right) \left(1-c_d\right)} \bigg)   \nonumber 
\end{align}
\begin{align}
    C(x) &= -\frac{(c_c+x) (1+x)(1-x)^2 e^{-\mathrm{i} (2\Psi _c+\Psi _d)}}{32  \left(1-c_c^2\right) \Gamma_1(x) \Gamma_2(x) } \\
    &\times\left(\sqrt{\left(1-c_b\right) \left(1+c_c\right)} e^{\mathrm{i} \Psi
   _c}-\sqrt{\left(1+c_b\right) \left(1-c_c\right)}\right) \left(\sqrt{\left(1+c_c\right) \left(1+c_b\right)} e^{\mathrm{i}
   \Psi _c}+\sqrt{\left(1-c_c\right) \left(1-c_b\right)}\right)^2 \nonumber \\
  &\times\left(\sqrt{\left(1-c_d\right) \left(1+c_c\right)} e^{\mathrm{i} \Psi
   _c}-\sqrt{\left(1+c_d\right) \left(1-c_c\right)} e^{i \Psi
   _d}\right) \left(\sqrt{\left(1+c_c\right) \left(1+c_d\right)} e^{\mathrm{i}
   \Psi _c}+\sqrt{\left(1-c_c\right) \left(1-c_d\right)} e^{i
   \Psi _d}\right)^2 \nonumber
\end{align}
\begin{align}
    D(x) &= \frac{(c_d+x) e^{-\mathrm{i} (\Psi _c+2\Psi _d)}}{32 (1+x) \left(1-c_d^2\right) \Gamma_2(x) \bigg(\sqrt{\left(1+c_d\right)
   \left(1-c_b\right)} -\sqrt{\left(1-c_d\right) \left(1+c_b\right)} e^{\mathrm{i} \Psi _d} \bigg) } \\
    &\times\left(\sqrt{\left(1-c_c\right) \left(1+c_d\right)} e^{\mathrm{i} \Psi
   _c}-\sqrt{\left(1+c_c\right) \left(1-c_d\right)}  e^{\mathrm{i} \Psi
   _d}\right) \left(\sqrt{\left(1+c_c\right) \left(1+c_d\right)} e^{\mathrm{i}
   \Psi _c}+\sqrt{\left(1-c_c\right) \left(1-c_d\right)}  e^{\mathrm{i} \Psi
   _d}\right)^2 \nonumber \\
   &\times \left((1+x)  \sqrt{\left(1+c_d\right)
   \left(1-c_b\right)}  + (1-x)
   \sqrt{\left(1-c_d\right) \left(1+c_b\right)} e^{\mathrm{i} \Psi _d}\right)^2 \nonumber \\
   &\times \bigg(  (1+x)
   \sqrt{\left(1+c_b\right) \left(1+c_d\right)} - (1-x) \sqrt{\left(1-c_b\right) \left(1-c_d\right)} e^{\mathrm{i} \Psi _d} \bigg)   \nonumber ,
\end{align}
where
\begin{align}
    \Gamma_1(x) &= (1+x)  \sqrt{\left(1+c_c\right)
   \left(1+c_b\right)} e^{\mathrm{i} \Psi _c} - (1-x) \sqrt{\left(1-c_c\right) \left(1-c_b\right)}  \; , \\
   \Gamma_2(x) &=  (1+x)  \sqrt{\left(1+c_c\right)
   \left(1+c_d\right)} e^{\mathrm{i} \Psi _c} - (1-x) \sqrt{\left(1-c_c\right) \left(1-c_d\right)} e^{\mathrm{i} \Psi _d} .
\end{align}
For certain parameter choices, the expressions above for $B(x)$, $C(x)$, $D(x)$ are ill-defined. This occurs in the following situations:
\begin{itemize}
    \item $c_b = \pm 1$ (i.e. $\gamma_{ab} = 0$ or $\gamma_{ab}=\pi$): in this case, the $g_b$ denominator in the integrand~\eqref{eq:defLambda} is independent of $\beta$, so there are no associated poles to be captured by the residue theorem. The same applies to  $g_c$ or $g_d$ when $c_c = \pm 1$ or $c_d = \pm 1$, respectively. 
    \item $\Psi_d = 0$ and $c_b=c_d$: here, the pole at $1/z_b^* = 1/z_d^*$ is actually a double pole.
\item $\Psi_c=0$ and $c_c \leq c_b$: in this case that there is a value of $x \in [-c_b; -c_c]$ for which the denominator $\Gamma_1(x)$, entering both $B(x)$ and $C(x)$ vanishes ( with the condition $c_b \leq c_c$ ensuring that the integration range in Eq.~\eqref{eq:residuesSimplePoles} actually contains the pole). Nevertheless, it can be checked that $B(x) + C(x)$ remains regular at this value of $x$, so that the total integral defining $\lambda_{abcd}$ in Eq.~\eqref{eq:defLambda} is still convergent. However, in the special case $\Psi_c=0$ and $c_c = c_b$, there remains a double pole in the integral.
\item $\Psi_c = \Psi_d$ and $c_c \leq c_d$: same issue as before, but for $\Gamma_2 (x)$.
\end{itemize}
We therefore exclude these special cases from the computation. The value of $\lambda_{abcd}$ can then be obtained by taking appropriate limits of the final expressions, which remain finite.

It remains to integrate these expressions with respect to $x = \cos \theta$.
Since the integrands are rational fractions of $x$, the indefinite integrals can be computed quite straightforwardly, albeit at the cost of producing very lengthy expressions. We find that the simplest expressions arise when manually performing the rational fraction decomposition of the integrands $B(x)$, $C(x)$, $D(x)$, instead of using the automated \textit{integrate} function in \textsc{Mathematica}. Furthermore, we have to pay attention to the fact that
\begin{equation}
     \int_\alpha^\beta \frac{1}{z} \, \mathrm{d}z = \log  \bigg( \frac{\beta}{\alpha} \bigg),
\end{equation}
which may be different from $\log \beta - \log \alpha$ depending on the arguments of the complex numbers $\alpha$ and $\beta$. 

The final expression of the 4PT correlator is obtained by summing over all residues, and is provided in the companion \textsc{Mathematica} notebook~\cite{mathematica}. Simplifying our expressions as much as possible, we find that it takes the remarkably simple form written in Eq.~\eqref{eq:lambdaabcd}, where the logarithmic factors only involve pairwise angles of pulsars. We have checked our results against direct numerical integration of Eq.~\eqref{eq:defLambda} for a few randomly chosen angles.

Let us comment on the validity of this result. \textit{A priori}, our calculation was fully general, but a few special cases had to be excluded from the computation, as explained above. In the final formula, these exclusions appear as nontrivial denominators or logarithmic arguments, so that a naive substitution of the corresponding parameter values leads to indeterminate expressions. However, the \textit{limit} of the 4PT function at these parameter values is of course finite and regular due to non-trivial cancellations. This is to be expected: after all, even the HD function itself $\mu(\gamma)$ is ill-defined at zero angular separation $\gamma=0$ (as it contains a term $\log(1-\cos \gamma)$), yet its limit is finite. To sum up, the parameter values for which the 4PT function is formally ill-defined but has a finite limit are:
\begin{itemize}
    \item $\gamma_{ab} = 0,\pi$, $\gamma_{ac} = 0,\pi$, $\gamma_{ad} = 0,\pi$,
    \item $\Psi_c=0$ and $\gamma_{ab} = \gamma_{ac}$,
    \item $\Psi_d = 0$ and $\gamma_{ab}=\gamma_{ad}$,
    \item $\Psi_c=\Psi_d$ and $\gamma_{ac}=\gamma_{ad}$.
\end{itemize}

\section{Four-point function in terms of real Fourier coefficients}\label{sec:realfourier}

In Sec.~\ref{sec:4pt}, we expanded the timing residuals induced by a SGWB generated by a population of SMBHBs in terms of a \textit{complex} Fourier series (see Eqs.~\eqref{eq:timingResidualSGWB} and \eqref{eq:Ccoeff}). However, in PTAs, the SGWB is usually modeled in a real Fourier basis, as shown by Eq.~\eqref{eq:sia}. The relation between the two basis yields
\begin{equation}
    \begin{cases}
        C_I^a = \frac{1}{2}(a_I^a - \mathrm{i} b_I^a)\\
        C_I^{a*} = \frac{1}{2}(a_I^a + \mathrm{i} b_I^a)
    \end{cases}\longrightarrow
    \begin{cases}
        a_I^a = C_I^a + C_I^{a*} \\
        b_I^a = \mathrm i(C_I^{a} - C_I^{a*}  )
    \end{cases}.
\end{equation}
With this substitution, it becomes quite easy to compute the quantities relevant to PTA data analysis in terms of the complex 4PT correlator described in Sec.~\ref{sec:4pt}. As an example, let us compute the following quantity:
\begin{equation}\label{eq:aaaa}
    \begin{split}
    \la a_I^a~ a_I^b ~a_I^c~ a_I^d\ra   &= \la (C_I^a + C_I^{a*})~(C_I^b + C_I^{b*})~(C_I^c + C_I^{c*})~(C_I^d + C_I^{d*})\ra_{\phi, \vec\Omega}   = \\
    &= \left(\la C_I^a~C_I^b~C_I^{c*}~C_I^{d*}\ra_{\phi, \vec\Omega}    + \la C_I^{a*}~C_I^{b*}~C_I^c~C_I^d\ra_{\phi, \vec\Omega}   \right) + \left(\la C_I^a~C_I^{b*}~C_I^c~C_I^{d*}\ra_{\phi, \vec\Omega}    + \la C_I^{a*}~C_I^b~C_I^{c*}~C_I^d\ra_{\phi, \vec\Omega}   \right) + \\
    & + \left(\la C_I^a~C_I^{b*}~C_I^{c*}~C_I^{d}\ra_{\phi, \vec\Omega}    + \la C_I^{a*}~C_I^b~C_I^{c}~C_I^{d*}\ra_{\phi, \vec\Omega}   \right) = \\
    &=\frac{1}{8}\mathcal{H}_4^I\eta_{abcd}~\text{Re}[\lambda_{abcd}] + \frac{1}{8}\left[(\mathcal{H}_2^I)^2 - \mathcal{H}_4^I\right]\left[\bar \mu(\gamma_{ab})\bar \mu(\gamma_{cd}) + \bar \mu(\gamma_{ad})\bar \mu(\gamma_{bc})\right] + (b \leftrightarrow c) + (c\leftrightarrow d).
    \end{split}
\end{equation}
where, analogously to Eq.~\eqref{eq:4ptCorr}, the average is performed over the intrinsic orbital phase $\phi$ and the solid angle $\Omega$.
In the previous equation, we kept only the terms whose average returns a non-vanishing quantity, along the lines of what explained in Sec.~\ref{sec:4pt}.
Notice that each term comes with its complex conjugate. Therefore, it is immediate to realize that the relevant quantity for PTAs is real --as explicitly demonstrated by the last line of Eq.~\eqref{eq:aaaa}-- even though the average of the $C_I^a$ can be complex.
We remind that, as shown in Eq.~\eqref{eq:4ptCorrN}, performing the average over the total number of sources in the model eliminates the term proportional to $\mathcal{H}_4^I$ in the square bracket in Eq.~\eqref{eq:aaaa}. \\
Analogously to Eqs.~\eqref{eq:4ptCorr} and \eqref{eq:4ptCorrN}, the 4PT of real Fourier coefficients contains a gaussian part, proportional to $(\mathcal{H}_2^I)^2$, and a purely non-gaussian component, proportional to $\mathcal{H}_4^I$.  \\
For illustration purposes, we also compute the quantities
\begin{equation}\label{eq:baaa}
    \begin{split}
    \la b_I^a~ a_I^b ~a_I^c~ a_I^d\ra   &= \la \mathrm{i}(C_I^a - C_I^{a*})~(C_I^b + C_I^{b*})~(C_I^c + C_I^{c*})~(C_I^d + C_I^{d*})\ra_{\phi,\vec\Omega}   = \\
    &= -\frac{1}{8}\mathcal{H}_4^I~\left\{\eta_{abcd}~\text{Im}[\lambda_{abcd}] + (b \leftrightarrow c) + (c \leftrightarrow d)\right\},\\
    \la a_I^a~ b_I^b ~b_I^c~ b_I^d\ra   &=- \la \mathrm{i}(C_I^a + C_I^{a*})~(C_I^b - C_I^{b*})~(C_I^c - C_I^{c*})~(C_I^d - C_I^{d*})\ra_{\phi,\vec\Omega}   = \\
    & =  \frac{1}{8}\mathcal{H}_4^I~\left\{\eta_{abcd}~\text{Im}[\lambda_{abcd}] + (b \leftrightarrow c) + (c \leftrightarrow d)\right\},
    \end{split}
\end{equation}
irrespectively of whether we perform or not the average over the total number of sources. Even in this case, it is immediate to realize that the correlators \eqref{eq:baaa} measured by PTAs is real, as expected.
In the usual gaussian case, the correlation $\la b_I^a a_I^b\ra  $ vanishes, as written in Eq.~\eqref{eq:sgwbcov}. In fact, because $\la F_+^a F_\times^b-F_\times^a F_+^b\ra_{\phi,\vec\Omega}   =0$ (see Ref.~\cite{Allen:2022dzg}, Appendix E),
\begin{equation}
    \la b_I^a a_I^b\ra   = \mathrm{i}\left(\la C_I^a C_I^{b*}\ra_{\phi,\vec\Omega}   - \la C_I^{a*} C_I^b\ra_{\phi,\vec\Omega}  \right) \propto \left(\la F_+^a F_+^b + F_\times^a F_\times^b \ra_{\phi,\vec\Omega} - \la F_+^a F_+^b + F_\times^a F_\times^b \ra_{\phi,\vec\Omega}  \right).
\end{equation}
Therefore, one might have expected the correlators in Eq.~\eqref{eq:baaa} to identically vanish as well. However, in that case, we have terms involving the average of products of four pattern functions, e.g. $\la F_+^a F_\times^b F_\times^c F_\times^d\ra_{\phi,\vec\Omega}   $. These products do not vanish, and enter with a different overall sign whether we consider $\la C_I^a~(C_I^b + C_I^{b*})~(C_I^c + C_I^{c*})~(C_I^d + C_I^{d*})\ra_{\phi,\vec\Omega}  $ or $\la C_I^{a*}~(C_I^b + C_I^{b*})~(C_I^c + C_I^{c*})~(C_I^d + C_I^{d*})\ra_{\phi,\vec\Omega}  $, for instance. Thus, the quantities~\eqref{eq:baaa} do not vanish. If the 2PT correlation acquires a nonzero imaginary part, as may occur due to
pulsar-term corrections or circular polarization, nonvanishing sine--cosine
covariances are generated and the second line of Eq.~\eqref{eq:sgwbcov} must be modified
accordingly \cite{Bernardo:2025dat}. The impact of pulsar-term corrections and circular
polarization on the 4PT correlators studied here has not yet been worked
out and is left for future investigation.\\
Finally, let us also compute
\begin{equation}\label{eq:aabb}
    \begin{split}
    \la a_I^a~ a_I^b ~b_I^c~ b_I^d\ra   &= - \la (C_I^a + C_I^{a*})~(C_I^b + C_I^{b*})~(C_I^c - C_I^{c*})~(C_I^d - C_I^{d*})\ra_{\phi,\vec\Omega}   = \\
    &= \frac{1}{8}\mathcal{H}_4^I~\left(\eta_{abcd}~\text{Re}[\lambda_{abcd}] + \eta_{abdc}~\text{Re}[\lambda_{abdc}] - \eta_{acbd}~\text{Re}[\lambda_{acbd}]\right) + \frac{1}{4}\left[(\mathcal{H}_2^I)^2 - \mathcal{H}_4^I\right]\left[\bar \mu(\gamma_{ab})\bar \mu(\gamma_{cd})\right].
    \end{split}
\end{equation}

Following the logic in Eqs.~\eqref{eq:aaaa} and \eqref{eq:baaa}, and using the results of Sec.~\ref{sec:4pt}, we can easily compute the connected 4PT correlator $\la w_I^a w_J^b w_K^c w_L^d\ra_c$ needed in Eq.~\eqref{eq:ngmarg} for any set of pulsars and GW frequencies. In fact, the only non-zero elements of the connected 4PT correlator will be given by Eqs.~\eqref{eq:aaaa} and \eqref{eq:baaa}, provided we remove the gaussian contribution proportional to $(\mathcal{H}_2^I)^2$. Moreover, performing the average over the fluctuations in the total number of sources show that the non-gaussianity is only given by the first terms in the r.h.s. of Eq.~\eqref{eq:aaaa} and \eqref{eq:aabb}, as detailed in Sec.~\ref{subsubsec:simp}.

\section{PTA marginalized likelihood}\label{sec:marg}
In this Appendix, we present a series of technical formulae that complement the discussion in Sec.~\ref{sec:pta}. First, we analyze the case of a gaussian prior distribution for the Fourier coefficients of the SGWB. Then, we extend the prior probability distribution by adding the contributions of the 4PT computed in Sec.~\ref{sec:4pt}.\\
We will mostly follow Appendices B-C of Ref.~\cite{Falxa:2025} and the treatment in Ref.~\cite{Kitaura_2012}, which we refer the reader to for further details.

\subsection{Gaussian case}\label{subsec:agauss}
As discussed in Sec.~\ref{sec:pta}, in the gaussian approximation, the Fourier coefficients are extracted from the prior in Eq.~\eqref{eq:prior}, which we recall here for convenience:
\begin{equation}\label{eq:aprior}
    \Pi_\text{G}(\vec{w}|\vec{\eta}) = \frac{\exp (-\frac{1}{2}\vec{w}^T \Phi^{-1} \vec{w})}{\sqrt{(2\pi)^W|\Phi|}},
\end{equation}
where $\Phi$ is the covariance matrix described by Eqs.~\eqref{eq:sgwbcov}-\eqref{eq:sgwbcov2}. The marginalized likelihood of Eq.~\eqref{eq:marglik}, then, will read:
\begin{equation}\label{eq:fi}
    \mathcal{L}^\text{G}_\text{marg}(\vec{\delta t}|\vec{\eta}) = \int~d\vec w~ \frac{\exp\left[-\frac{1}{2}\left(\vec{\delta t} - F\vec{w}\right)^T \mathcal{N}^{-1} \left(\vec{\delta t} - F\vec{w}\right)\right]}{\sqrt{(2\pi)^N|\mathcal{N}|}} \frac{\exp (-\frac{1}{2}\vec{w}^T \Phi^{-1} \vec{w})}{\sqrt{(2\pi)^W|\Phi|}},
\end{equation}
where we have used Eq.~\eqref{eq:lik} for the likelihood of timing residuals. Following the procedure detailed in Appendix B of Ref.~\cite{Falxa:2025}, Eq.~\eqref{eq:fi} becomes:
\begin{equation}
    \mathcal{L}^\text{G}_\text{marg}(\vec{\delta t}|\vec{\eta}) = \int~d\vec w~ \frac{\exp\left[-\frac{1}{2}
    \left(\vec{\delta t}\,^T\mathcal{N}^{-1}\vec{\delta t} + \vec{w}^T(F^T\mathcal{N}^{-1}F + \Phi^{-1})\vec w - 2~\vec w^TF^T\mathcal{N}^{-1}\vec{\delta t}\right)
    \right]
    }
    {\sqrt{(2\pi)^N|\mathcal{N}|}\sqrt{(2\pi)^W|\Phi|}}.
\end{equation}
Defining $\Sigma = F^T\mathcal{N}^{-1}F + \Phi^{-1}$ and completing the squares, we can rewrite the previous integral as:
\begin{equation}\label{eq:ii}
    \begin{split}
    \mathcal{L}^\text{G}_\text{marg}(\vec{\delta t}|\vec{\eta}) = &\frac{\exp\left[-\frac{1}{2}\vec{\delta t}\,^T\left( \mathcal{N}^{-1} - \mathcal{N}^{-1} F\,\Sigma^{-1}F^T\mathcal{N}^{-1}\right)\vec{\delta t}\right]}{\sqrt{(2\pi)^N|\mathcal{N}|}\sqrt{(2\pi)^W|\Phi|}}\times\\
    &\times\int~d\vec w~ \exp\left[-\frac{1}{2}
    \left(\vec{w} - \Sigma^{-1}F^T\mathcal{N}^{-1}\vec{\delta t} \right)^T\Sigma\left(\vec{w} - \Sigma^{-1}F^T\mathcal{N}^{-1}\vec{\delta t} \right)
    \right].
    \end{split}
\end{equation}
Performing the integral yields:
\begin{equation}
    \mathcal{L}^\text{G}_\text{marg}(\vec{\delta t}|\vec{\eta}) = \frac{\exp\left[-\frac{1}{2}\vec{\delta t}\,^T\left( \mathcal{N}^{-1} - \mathcal{N}^{-1} F\,\Sigma^{-1}F^T\mathcal{N}^{-1}\right)\vec{\delta t}\right]}{\sqrt{(2\pi)^N|\mathcal{N}||\Phi||\Sigma|}}.
\end{equation}
This is the marginalized likelihood commonly used in PTA data analysis, derived assuming that the Fourier coefficients of the SGWB are distributed according to a gaussian prior. Sometimes, the exponent is conveniently rewritten making use of the Woodbury lemma~\cite{Woodbury:1950} $[\mathcal{N} + F\Phi F^T]^{-1} =\mathcal{N}^{-1} - \mathcal{N}^{-1} F\,\Sigma^{-1}F^T\mathcal{N}^{-1}$.

\subsection{Non gaussian case}\label{sec:ang}
As already mentioned in Sec.~\ref{subsec:nglik}, accounting for non-gaussianity means that the gaussian prior in Eq.~\eqref{eq:aprior} will not encode the full statistics of the Fourier coefficients describing the SGWB. In this Section, we aim at understanding how the 4PT described in Sec.~\ref{sec:4pt} enters the PTA inference pipeline. To do so, we derive a marginalized likelihood, following the same conceptual steps detailed in Appendix~\ref{subsec:agauss}. A convenient way to introduce non-gaussianity is by using the multivariate Edgeworth expansion~\cite{Kitaura_2012} :
\begin{equation}\label{eq:angprior}
    \begin{split}
    \Pi(\vec w|\vec{\eta}) = \Pi_\text{G}(\vec w|\vec \eta)\times\left[1 + \frac{1}{4!}\sum_{\{a'\}}\sum_{\{I'\}}\sum_{\{a\}}\sum_{\{I\}}  (\kappa_4)^{a'b'c'd'}_{I'J'K'L'} (\Phi^{-1/2})^{aa'}_{II'}(\Phi^{-1/2})^{bb'}_{JJ'}(\Phi^{-1/2})^{cc'}_{KK'}(\Phi^{-1/2})^{dd'}_{LL'} H^{abcd}_{IJKL}(\vec{\nu}) \right],
    \end{split}
\end{equation}
where
\begin{equation}
\begin{split}
    (\kappa_4)^{abcd}_{IJKL} \equiv \la w^{a}_{I}w^{b}_{J}w^{c}_{K}w^{d}_{L}\ra  _c  &=\la w^{a}_{I}w^{b}_{J}w^{c}_{K}w^{d}_{L}\ra   - \la w^{a}_{I}w^{b}_{J}\ra  \la w^{c}_{K}w^{d}_{L}\ra -\la w^{a}_{I}w^{c}_{K}\ra  \la w^{b}_{J}w^{d}_{L}\ra  - \la w^{a}_{I}w^{d}_{L}\ra  \la w^{b}_{J}w^{c}_{K}\ra  .
\end{split}
\end{equation}
is the \textit{connected} 4PT correlator of the real Fourier coefficients and we conveniently group the indices as $\{a'\} =\{a',b',c',d'\}$, for instance. The Edgeworth series is an asymptotic expansion and can exhibit pathological behavior in the tails of the distribution, such as yielding negative values for the probability density. For this reason, we restrict our analysis to the regime in which non-gaussian corrections are perturbatively small compared to the underlying gaussian contribution (see discussion below Eq.~\eqref{eq:nuI}).\\
In principle, Eq.~\eqref{eq:angprior} should include also contributions arising from the 3PT correlator of Fourier coefficients. However, as we thoroughly discussed in Sec.~\ref{sec:bisp}, the 3PT correlator is vanishing, due to the superposition of the uncorrelated $\phi_I^i$ phases characterizing SMBHB systems.\\
Eq.~\eqref{eq:angprior} involves a sum over the Hermite polynomials, defined in terms of the unit-variance variable $\vec\nu$, whose components are determined by 
\begin{equation}\label{eq:anu}
    \nu^a_I = \left(\Phi^{-1/2} \vec{w} \right)^{a}_{I},
\end{equation}
presented in Eq.~\eqref{eq:nuI} and reported here for convenience.~\footnote{Notice that, since $\Phi^{-1}$ is the inverse of a covariance matrix, it is symmetric and positive definite. Therefore, the matrix square root $\Phi^{-1/2}$ appearing in Eqs.~\eqref{eq:angprior} and \eqref{eq:anu} is well defined and can be written as $\Phi^{-1/2} \equiv O \left(\Phi^{-1}_{\mathrm{diag}}\right)^{1/2} O^{T}$, where $O$ is the orthogonal matrix that diagonalizes $\Phi^{-1}$ and $\Phi^{-1}_{\mathrm{diag}}$ is the diagonal matrix of its (strictly positive) eigenvalues. Anyway, the final result will only depend on factors involving $\Phi^{-1}$. Therefore, we can treat this as a \textit{formal} step, without worrying too much about the details. }
The Hermite polynomials are then~\cite{Kitaura_2012}:
\begin{equation}\label{eq:4herm}
\begin{split}
    H_{IJKL}^{abcd} (\vec\nu) = &~\nu_I^a\nu^b_J\nu^c_K\nu^d_L - \left(\nu_I^a\nu_J^b~ \delta^{cd}_{KL} + \nu_I^a\nu_K^c ~\delta^{bd}_{JL} + \nu_I^a\nu_L^d~ \delta^{bc}_{JK} + \nu_J^b\nu_K^c ~\delta^{ad}_{IL} + \nu_J^b\nu_L^d~ \delta^{ac}_{IK} + \nu_K^c\nu_L^d ~\delta^{ab}_{IJ}\right) \\
    +& ~\delta^{ab}_{IJ} ~\delta^{cd}_{KL} + \delta^{ac}_{IK}~ \delta^{bd}_{JL} + \delta^{ad}_{IL}~ \delta^{bc}_{JK}.
\end{split}
\end{equation}
As a qualitative check, we note that in the one-dimensional case the fourth Hermite polynomial is $h_4(\nu) = \nu^4 - 6\nu^2 + 3$, which essentially preserves the same structure as Eq.~\eqref{eq:4herm}.
Plugging Eq.~\eqref{eq:4herm}, together with the substitution~\eqref{eq:anu}, in Eq.~\eqref{eq:angprior}, we find
\begin{equation}\label{eq:apri}
    \begin{split}
    \Pi(\vec w|\vec{\eta}) = \Pi_\text{G}(\vec w|\vec \eta)\times&\left[1 + \frac{1}{4!}\sum_{\{a\}}\sum_{\{I\}} (\kappa_4)^{abcd}_{IJKL}\times \left((\Phi^{-1}\vec{w})^{a}_{I}(\Phi^{-1}\vec{w})^{b}_{J}(\Phi^{-1}\vec{w})^{c}_{K}(\Phi^{-1}\vec{w})^{d}_{L}~  - \right.\right.\\
    &-\left.\left. (\Phi^{-1}\vec{w})^{a}_{I}(\Phi^{-1})^{cd}_{KL}(\Phi^{-1}\vec{w})^{b}_{J} -  (\Phi^{-1}\vec{w})^{a}_{I}(\Phi^{-1})^{bd}_{JL}(\Phi^{-1}\vec{w})^{c}_{K} -\right.\right.\\
    &- \left.\left.  (\Phi^{-1}\vec{w})^{a}_{I}(\Phi^{-1}\vec{w})^{d}_{L}(\Phi^{-1})^{bc}_{JK} - (\Phi^{-1}\vec w)^{b}_{J}(\Phi^{-1}\vec w)^{c}_{K}(\Phi^{-1})^{ad}_{IL} -\right.\right.\\
    &-\left.\left.  (\Phi^{-1}\vec w)^{b}_{J}(\Phi^{-1}\vec w)^{d}_{L}(\Phi^{-1})^{ac}_{IK} -  (\Phi^{-1}\vec w)^{c}_{K}(\Phi^{-1}\vec w)^{d}_{L}(\Phi^{-1})^{ab}_{IJ} +\right.\right.\\
    &+ \left.\left.(\Phi^{-1})^{ab}_{IJ}(\Phi^{-1})^{cd}_{KL} + (\Phi^{-1})^{ac}_{IK}(\Phi^{-1})^{bd}_{JL} + (\Phi^{-1})^{ad}_{IL}(\Phi^{-1})^{bc}_{JK}\right)\right].
    \end{split}
\end{equation}

In order to arrive to this expression, we used the relation
\begin{equation}
    \sum_a \sum_I\left(\Phi^{-1/2}\right)^{a'a}_{I'I}\left(\Phi^{-1/2}\right)^{aa''}_{II''} = \left(\Phi^{-1}\right)^{a'a''}_{I'I''}.
\end{equation}
We can simplify the expression in Eq.~\eqref{eq:apri} by noticing some properties. First, in this work we are assuming different frequency bins to be uncorrelated.~\footnote{As discussed in Sec.~\ref{sec:pta}, a finite observation time leads to inter-frequency correlations~\cite{Crisostomi:2025vue}, which can be consistently incorporated through the use of window functions~\cite{Lamb:2025niq}. For simplicity, we neglect these correlations in the present analysis. Nevertheless, including them poses no conceptual difficulty: one simply needs to consistently retain all indices in Eq.~\eqref{eq:apri}, without invoking the $\delta$-function simplification employed later in Eq.~\eqref{eq:ad}. } That is to say
\begin{equation}\label{eq:ad}
    \left(\Phi^{-1}\right)^{ab}_{IJ} \propto \delta_{IJ}.
\end{equation}
Secondly, the connected 4PT correlation function $(\kappa_4)^{abcd}_{IJKL}$ is completely symmetric under any switch involving the couples of indices $(a,I), (b,J), (c,K)$ and $ (d,L)$. With these two properties, Eq.~\eqref{eq:apri} reduces to the much more readable form:
\begin{equation}\label{eq:aprired}
    \begin{split}
    \Pi(\vec w|\vec{\eta}) = \Pi_\text{G}(\vec w|\vec \eta)\times&\left[1 + \frac{1}{4!}\sum_{abcd}\sum_{IJKL} (\kappa_4)^{abcd}_{IJKL}\left((\Phi^{-1}\vec w)^{a}_{I}(\Phi^{-1}\vec w)^{b}_{J}(\Phi^{-1}\vec w )^{c}_{K}(\Phi^{-1}\vec{w})^{d}_{L}  -\right.\right.\\
    &-\left.\left. 6 (\Phi^{-1}\vec w)^{a}_{I}(\Phi^{-1}\vec w)^{b}_{J}(\Phi^{-1})^{cd}_{KL} +3(\Phi^{-1})^{ab}_{IJ}(\Phi^{-1})^{cd}_{KL}\right)\right],
    \end{split}
\end{equation}
where now 
\begin{equation}\label{eq:simpphi}
    \left(\Phi^{-1}\vec{w}\right)^{a}_I = \sum_{a'} \left(\Phi^{-1}\right)^{aa'}_{II} w^{a'}_I.
\end{equation}
In analogy to the gaussian case, we now turn to the determination of the marginalized likelihood, including the non-gaussian contribution encoded by the 4PT correlator. Thus, we have to compute the following integral:
\begin{equation}
    \mathcal{L}_\text{marg}(\vec{\delta t}|\vec{\eta}) = \int~d\vec w~ \frac{\exp\left[-\frac{1}{2}\left(\vec{\delta t} - F\vec{w}\right)^T \mathcal{N}^{-1} \left(\vec{\delta t} - F\vec{w}\right)\right]}{\sqrt{(2\pi)^N|\mathcal{N}|}} \Pi(\vec w|\vec \eta),
\end{equation}
where the likelihood is the same of the gaussian case, while the prior encoding non-gaussianity is given by Eq.~\eqref{eq:aprired}
Following the same procedure as in Sec.~\ref{subsec:agauss}, we arrive to:
\begin{equation}\label{eq:alikng}
    \begin{split}
    \mathcal{L}_\text{marg}(\vec{\delta t}|\vec{\eta}) = &\frac{\exp\left[-\frac{1}{2}\vec{\delta t}\,^T\left( \mathcal{N}^{-1} - \mathcal{N}^{-1} F\,\Sigma^{-1}F^T\mathcal{N}^{-1}\right)\vec{\delta t}\right]}{\sqrt{(2\pi)^N|\mathcal{N}||\Phi||\Sigma|}}\times\\
    &\times\left[1 + \frac{1}{4!}\sum_{\{a\}}\sum_{\{I\}} (\kappa_4)^{abcd}_{IJKL}\left(\sum_{\{a'\}}(\Phi^{-1})^{aa'}_{II}(\Phi^{-1})^{bb'}_{JJ}(\Phi^{-1})^{cc'}_{KK}(\Phi^{-1})^{dd'}_{LL}~ E[w_{I}^{a'} w_{J}^{b'} w_{K}^{c'} w_{L}^{d'}]-  \right.\right.\\
    &-\left.\left. 6\sum_{\{a'\}} (\Phi^{-1})^{aa'}_{II}(\Phi^{-1})^{bb'}_{JJ}(\Phi^{-1})^{cd}_{KL}~E[w_{I}^{a'} w_{J}^{b'}] +3(\Phi^{-1})^{ab}_{IJ}(\Phi^{-1})^{cd}_{KL}\right)\right],
    \end{split}
\end{equation}
where
\begin{equation}
    E[w^{a}_{I}...] \equiv \sqrt{\frac{|\Sigma|}{(2\pi)^W}} \int~d\vec w~ \exp\left[-\frac{1}{2}
    \left(\vec{w} - \hat w \right)^T\Sigma\left(\vec{w} - \hat w \right)\right]w^{a}_{I}... ,
\end{equation}
with $\hat w$ defined in  Eq.~\eqref{eq:sub}.
The Isserlis's theorem~\cite{Isserlis:1918} (or Wick's theorem~\cite{Wick:1950}) allows us to explicitly compute the expectation values of Eq.~\eqref{eq:alikng} as:
\begin{equation}
    \begin{split}
        E[w_{I}^{a} w_{J}^{b} w_{K}^{c} w_{L}^{d}] = &\hat w_{I}^{a} \hat w_{J}^{b}\hat w_{K}^{c}\hat w_{L}^{d} + \hat w_{I}^{a} \hat w_{J}^{b} \left(\Sigma^{-1}\right)_{KL}^{c d} + \hat w_{I}^{a} \hat w_{K}^{c} \left(\Sigma^{-1}\right)_{JL}^{b d}+ \\
        + &\hat w_{I}^{a} \hat w_{L}^{d} \left(\Sigma^{-1}\right)_{JK}^{bc} + \hat w_{J}^{b} \hat w_{K}^{c} \left(\Sigma^{-1}\right)_{IL}^{a d} + \hat w_{J}^{b} \hat w_{L}^{d} \left(\Sigma^{-1}\right)_{IK}^{a c} + \\
        +&\hat w_{K}^{c} \hat w_{L}^{d} \left(\Sigma^{-1}\right)_{IJ}^{a b} + \left(\Sigma^{-1}\right)_{IJ}^{a b}\left(\Sigma^{-1}\right)_{KL}^{c d} + \left(\Sigma^{-1}\right)_{IL}^{a d} \left(\Sigma^{-1}\right)_{JK}^{bc} + \\
        +&\left(\Sigma^{-1}\right)_{IK}^{a c}\left(\Sigma^{-1}\right)_{JL}^{b d} \\
        E[w_{I}^{a} w_{J}^{b}] = &\left(\Sigma^{-1}\right)_{IJ}^{a b} + \hat w_{I}^{a} \hat w_{J}^{b}.
    \end{split}
\end{equation}
By plugging these quantities in Eq.~\eqref{eq:alikng}, and recalling the symmetry of the 4PT correlation function, we finally find:
\begin{equation}\label{eq:alikngfin}
    \begin{split}
    &\mathcal{L}_\text{marg}(\vec{\delta t}|\vec{\eta}) = \mathcal{L}^\text{G}_\text{marg}(\vec{\delta t}|\vec{\eta}) \left\{1 + \frac{1}{4!}\sum_{\{a\}}\sum_{\{I\}} (\kappa_4)^{abcd}_{IJKL}\left[(\Phi^{-1}\hat w)^{a}_{I}(\Phi^{-1}\hat{w})^{b}_{J}(\Phi^{-1}\hat{w})^{c}_{K}(\Phi^{-1}\hat{w})^{d}_{L} \right.\right.\\
    &+\left.\left.~6~\left((\Phi^{-1}\hat w)_{I}^{a}(\Phi^{-1}\hat w)_{J}^{b} \left(\Phi^{-1}\Sigma^{-1}\Phi^{-1}\right)_{KL}^{cd} -\left(\Phi^{-1}\Sigma^{-1}\Phi^{-1}\right)_{IJ}^{ab}(\Phi^{-1})^{cd}_{KL} - (\Phi^{-1}\hat w)_{I}^{a} (\Phi^{-1}\hat w)_{J}^{b}(\Phi^{-1})^{cd}_{KL}\right)  \right.\right.\\
    &+\left.\left.~3~\left(\left(\Phi^{-1}\Sigma^{-1}\Phi^{-1}\right)_{IJ}^{a b}\left(\Phi^{-1}\Sigma^{-1}\Phi^{-1}\right)_{KL}^{c d} + (\Phi^{-1})^{ab}_{IJ}(\Phi^{-1})^{cd}_{KL}\right)\right]\right\}. 
    \end{split}
\end{equation}
where $(\Phi^{-1}\hat w)^{a}_{I}$ is given by Eq.~\eqref{eq:simpphi} and
\begin{equation}
    \left(\Phi^{-1}\Sigma^{-1}\Phi^{-1}\right)_{IJ}^{ab} = \sum_{\{a'\}} \left(\Phi^{-1}\right)^{aa'}_{II}\left(\Sigma^{-1}\right)^{a'b'}_{IJ}\left(\Phi^{-1}\right)^{b'b}_{JJ},
\end{equation}
thanks to the structure in Eq.~\eqref{eq:ad}.
This is the most general expression of a marginalized likelihood that includes non-gaussian contributions coming from the fourth order statistics.  \\
In theory, replacing the gaussian marginalized likelihood in Eq.~\eqref{eq:ii} with the expression in Eq.~\eqref{eq:alikngfin} in the PTA pipeline allows us to account for non-gaussian features of the SGWB. In practice, the implementation of the marginalized likelihood in Eq.~\eqref{eq:alikngfin} can be extremely expensive, as it requires knowing the explicit inverse of $\Sigma$. In the gaussian case, this cost is avoided by using a Cholesky-based linear solver, which computes directly the quantity $ \hat w$, without ever forming $\Sigma^{-1}$ explicitly~\cite{Falxa:2025}. 
While we do not attempt at proposing a practical and computationally-friendly method for the actual implementation of Eq.~\eqref{eq:alikngfin}, we note a couple of points that might go in the direction of reducing the computational burden.\\

Firstly, as shown explicitly in Sec.~\ref{sec:4pt}, the \textit{connected} part of the 4PT correlator only contains terms evaluated at the same frequency. These include all the terms of the form:
\begin{equation}
    \la a_I^a a_I^b a_I^c a_I^d\ra, \quad \la b_I^a b_I^b b_I^c b_I^d\ra, \quad \la a_I^a b_I^b b_I^c b_I^d\ra + \text{perm.}, \quad \la b_I^a a_I^b a_I^c a_I^d\ra + \text{perm.}, \quad \la b_I^a b_I^b a_I^c a_I^d\ra + \text{perm.},
\end{equation}
where $a_I^a, b_I^a$ are the coefficients of the real Fourier basis for pulsar $a$ (see Eq.~\eqref{eq:sia}). This observation drastically reduces the elements involved in the $\{I\}$ sum in Eq.~\eqref{eq:alikngfin}. 
Moreover, as already mentioned in the discussion below Eq.~\eqref{eq:nuI}, our analysis is expected to be most relevant in the intermediate frequency regime (roughly around $10^{-8}~\text{Hz}$), where the number of contributing SMBHBs remains substantial but not overwhelmingly large. With this in mind, it may be possible to further reduce the terms involved in the $\{I\}$ sum, selecting only the ones that are expected to yield the largest contribution. 

Secondly, most of the terms can be precomputed. In fact, it is easy to realize that
\begin{equation}
    \left(\Phi^{-1}\right)^{ab}_{II} = \frac{T}{S(f_I)} \Psi^{ab} ,
\end{equation}
where $\Psi^{ab}$ depends only on the HD correlations between different pairs of pulsars. In contrast, the dependence on the frequency, labeled by the index $I$, can be entirely factorized in the prefactor. While the hyperparameters defining the power spectrum $S(f_I)$, such as $(A_\text{gw}, \gamma_\text{gw})$ in Eq.~\eqref{eq:sf}, should ideally be treated as free parameters and sampled over in a PTA search, the matrix $\Psi$ can be easily precomputed, as the relative orientation of the pulsars is assumed fixed throughout the entire duration of the PTA experiment. \\
Moreover, Sec.~\ref{sec:4pt} and Appendix~\ref{sec:realfourier} demonstrate that the frequency dependence of the connected 4PT can be factorized as well. In fact, we have 
\begin{equation}\label{eq:akappa}
    \sum_{\{I\}}^{2N_f}\left(\kappa_4\right)_{IJKL}^{abcd} = \sum_G^{N_f}\sum_{\{n\} =0}^1 \left(\kappa_4\right)_{G+n, G+m, G+r, G+s}^{abcd} = \sum_{G}^{N_f} \frac{1}{8}\mathcal{H}_4^G\sum_{\{n\} =0}^1 \left(\tilde \kappa_4\right)_{n m r s}^{abcd},
\end{equation}
where the factors $\left(\tilde \kappa_4\right)_{n m r s}^{abcd}$ depend only on the relative orientations of the pulsars in the set $\{a,b,c,d\}$ and can therefore be precomputed using the method described in Appendix~\ref{sec:realfourier}.
We caution the reader to carefully track the notation in Eq.~\eqref{eq:akappa}.
The indices $\{I\}$ label real Fourier coefficients and therefore run up to $2N_f$, since each of the $N_f$ frequency bins is associated with two real Fourier modes. By contrast, the index $G$ labels the frequency bins themselves and thus ranges only up to $N_f$. The associated multiplicity is accounted for by the indices $\{n\}$, which select the specific real Fourier coefficients within each frequency bin.
The factor $\mathcal{H}_4^G$ encodes the frequency-dependent contribution at the frequency bin $G$.
As shown explicitly in Sec.~\ref{sec:4pt} and Appendix~\ref{sec:realfourier}, the factor $\mathcal{H}_4^G$ characterizes the frequency-dependent amplitude of non-gaussianity at frequency $f_G$, and should be sampled over as a free parameter in a Bayesian PTA analysis. \\
In light of Eq.~\eqref{eq:akappa}, it is instructive to estimate the number of distinct correlators that enter the sum in Eq.~\eqref{eq:alikng},  as a way to gauge the computational burden. Firstly, Eq.~\eqref{eq:akappa} shows that the frequency dependence of $\left(\kappa_4\right)_{IJKL}^{abcd}$ is fully factorized into the  $\mathcal{H}_4^G$ terms. Thus, the number of elements  $\left(\tilde \kappa_4\right)_{nmrs}^{abcd}$ that need to be computed for the sum  does not depend on the number of frequency bins considered. Below, we summarize the types of elements contained in $\left(\tilde\kappa_4\right)_{nmrs}^{abcd}$ along with their associated multiplicities, indicated with the symbol $N[...]$:
\begin{equation}\label{eq:cases}
\begin{cases}
     &N[\la a^a a^b a^c a^d\ra] = N[\la b^a b^b b^c b^d\ra]=\binom{N_p + 3}{4} \xrightarrow{N_p\gg1}\frac{N_p^4}{4!}\\
     &N[\la a^a a^b a^c b^d\ra]  = N[\la a^a b^b b^c b^d\ra]  =  N_p\binom{N_p + 2}{3} \xrightarrow{N_p\gg1} \frac{N_p^4}{3!}\\
      &N[\la a^a a^b b^c b^d\ra] = \binom{N_p + 1}{2}^2\xrightarrow{N_p\gg1} \frac{N_p^4}{4},\\
      & 
\end{cases}
\end{equation}
where we have used the fact that, for instance, $\la a^a a^b a^c b^d\ra$ is the same as $\la b^a a^b a^c a^d \ra$ up to indices redefinitions. Moreover, we have omitted the $\{I\}$ indices, since the result is completely independent on the specific frequency . Notice, however, that all the Fourier coefficients must be taken at the same frequency, otherwise the correlators vanish. For illustrative purposes, we also show the large-$N_p$ in Eq.~\eqref{eq:cases}, which is relevant as a quick estimate for current and future datasets. \\
Eq.~\eqref{eq:cases} immediately shows that the number of non-vanishing coefficients scales as $\mathcal{O}(N_p^4)$, implying a rapid growth with the total number of pulsars. Although the functional forms of these correlators have been computed in this work and remain fixed for every likelihood evaluation, their large number can significantly slow down computation. While addressing this optimization is beyond the scope of this paper, we note a possible workaround below Eq.~\eqref{eq:intre}, and discuss the issue further in Sec.~\ref{sec:concl}.

We now seek a simplified functional form that can be readily implemented in PTA data-analysis pipelines. Firstly, we rewrite Eq.~\eqref{eq:2ptCorr} as
\begin{equation}
 \la \mathcal{C}_I^a \mathcal{C}_{-I}^{b}\ra_{\phi, \vec\Omega} = \frac{1}{4}\mathcal{H}_2^I\bar \mu(\gamma_{ab}).    
\end{equation}
Expressed in terms of real Fourier coefficients (see Appendix~\ref{sec:realfourier}), this becomes
\begin{equation}
    \la a_I^a a_I^b \ra = \frac{1}{2}\mathcal{H}_2^I \bar \mu(\gamma_{ab}).
\end{equation}
By comparison with Eq.~\eqref{eq:sgwbcov}, we immediately identify
\begin{equation}
    \mathcal{H}_2^I = 2\frac{S(f_I)}{T}.
\end{equation}
On dimensional grounds, we can then assume 
\begin{equation}\label{eq:intre}
    \mathcal{H}_4^I = \epsilon_I(\mathcal{H}_2^I)^2 = 4~\epsilon_I \left(\frac{S(f_I)}{T}\right)^2,
\end{equation}
where the dimensionless coefficient $\epsilon_I$ quantifies the strength of the non-gaussian contribution relative to the (the square of) the gaussian one. In the limiting case of a single source in the $I-$th frequency bin, one expects $\epsilon_I = 1$, whereas $\epsilon_I\ll 1$ when many SMBHBs contribute to the bin. This formulation naturally promotes the set $\{\epsilon_I\}$ to parameters to be inferred in the analysis, with a naive prior range $\epsilon_I \in [0,1]$. We defer the exploration of alternative parametrizations of $\mathcal{H}_4^I$ to future work.\\

From the previous discussion, it might seem that we could write the sums in Eq.~\eqref{eq:alikngfin} as $\left(\sum_I^{N_f}X(f_I)\right) B$, where the sum schematically collects frequency-dependent term, while $B$ is a quantity depending only on the relative orientations of the pulsars. Unfortunately, this is not the case: in fact, the frequency dependent factors contained within $\Phi^{-1}$ enter the quantities $\Sigma^{-1} $ and $\hat w$ non-trivially. Therefore, the total number of non-zero terms in the sum in Eq.~\eqref{eq:alikng} is actually $\mathcal{O}(N_p^4 \times N_f)$. However, the simplifications outlined above may still help to consistently reduce the computational burden of Eq.~\eqref{eq:alikngfin}. 

Finally, let us comment on another assumption that might help implementing Eq.~\eqref{eq:alikngfin} in the PTA pipeline. As mentioned above, the hyperparameters defining the power spectrum $S(f_I)$, such as $(A_\text{gw}, \gamma_\text{gw})$ in Eq.~\eqref{eq:sf}, should ideally be treated as free parameters and sampled over in a PTA search. This means that the quantities $\hat w$ and $\Sigma^{-1}$ in Eq.~\eqref{eq:alikngfin} should be recomputed at every likelihood evaluation, which will definitely slow down the inference. \\
As a first approximation, however, we can fix the hyperparameters in the non-gaussian sum to the best-fit values obtained from the gaussian analysis. In other words, while we leave these parameters free in the term outside the curly brackets in Eq.~\eqref{eq:alikngfin}, we fix them within the sums at the gaussian best-fit values determined in a previous run. This way, the only remaining free parameters inside the sums are the $\epsilon_I$ introduced in Eq.~\eqref{eq:intre}, which quantify the strength of non-gaussianity at each frequency $f_I$, while the other contributions can be precomputed once. Therefore, even though the sum in Eq.~\eqref{eq:alikng} involves $\mathcal{O}(N_p^4 \times N_f)$ terms, they can be precomputed prior to the run, making the actual likelihood evaluation fast.\\
Since we are assuming that the non-gaussianity is relatively small compared to the gaussian background (otherwise the Edgeworth expansion in Eq.~\eqref{eq:angprior} would break down), this procedure is conceptually similar to perturbation theory: morally speaking, we are allowed to use the zeroth-order values of the coefficients in the first-order corrections, as any discrepancy would only contribute at higher order. \\
Of course, this work does not aim to provide a practical or computationally efficient implementation of the likelihood in Eq.~\eqref{eq:alikngfin} within the PTA pipeline. Accordingly, we leave the question of the optimal strategy for future investigation.

\bibliography{biblio}

\end{document}